\documentstyle[aps,prb,epsfig]{revtex}
\begin{document}
\twocolumn 
\wideabs{ 
\title{Plasmon excitations and 1D -
2D dimensional crossover in quantum crossbars} 
\author{I. Kuzmenko, S.
Gredeskul, K. Kikoin, Y. Avishai} 
\address{Department of Physics,
Ben-Gurion University of the Negev, Beer-Sheva} 
\date{\today}
\maketitle
\begin{abstract}
Spectrum of boson fields and two-point correlators are analyzed in
quantum crossbars (QCBs, a superlattice formed by $m$ crossed
interacting arrays of quantum wires), with short range inter-wire
capacitive interaction.  Spectral and correlation properties of double
($m=2$) and triple ($m-3$) QCBs are studied.  It is shown that the
standard bosonization procedure is valid, and the system behaves as a
sliding Luttinger liquid in the infrared limit, but the high frequency
spectral and correlation characteristics have either 1D or 2D nature
depending on the direction of the wave vector in the 2D elementary
cell of reciprocal lattice.  As a result, the crossover from 1D to 2D
regime may be experimentally observed. It manifests itself as
appearance of additional peaks of optical absorption, non-zero
transverse space correlators and periodic energy transfer between
arrays ("Rabi oscillations").
\end{abstract}
} \vspace{\baselineskip}
\section{Introduction}\label{sec:Intro}
The behavior of electrons in arrays of $1D$ quantum wires was
recognized a challenging problem soon after the consistent theory
of elementary excitations and correlations in a Luttinger liquid (LL)
of interacting electrons in one dimension was formulated (see
\cite{Voit} for a review).  One of fascinating targets
is a search for LL features in higher dimensions
\cite{Anders}.  Although the Fermi liquid state seems to be rather
robust for $D>1$, the possible way to retain some $1D$ excitation
modes in $2D$ and even $3D$ systems is to consider highly anisotropic
objects, in which the electron motion is spatially confined in major
part of the real space (e.g., it is confined to separate linear
regions by potential relief).  One may hope that in this case weak
enough interaction does not violate the generic long-wave properties
of the LL state.  Arrays of interacting quantum wires may be formed in
organic materials and in striped phases of doped transition metal
oxides.  Artificially fabricated structures with controllable
configurations of arrays and variable interactions are available now
due to recent achievements in nanotechnology (see, e.g.,
Refs.\onlinecite{Rueckes,Dai}).\\

The simplest $2D$ ensemble of $1D$ nanoobjects is an array of 
parallel quantum wires.  The conventional LL regime in a single $1D$ 
quantum wire is characterised by bosonic fields describing charge and 
spin modes.  We confine our discussion to the charge sector (LL in the 
spin-gapped phase).  The Hamiltonian of an isolated quantum wire may 
then be represented in a canonical form
\begin{equation}
H = \frac{\hbar v}{2}\int\limits_{-L/2}^{L/2} {dx}
\left\{g{\pi}^{2}(x)+
\frac{1}{g}({\partial}_{x}\theta^2(x))\right\}.
\label{D1}
\end{equation}
Here $L$ is the wire length, $v$ is the Fermi velocity, $\theta,\pi$
are the conventional canonically conjugated boson fields and $g$ is the
dimensionless parameter which describes the strength of the interaction
within the chain (see, e.g., \cite{Voit,Delft}).
The interwire interaction may transform the LL state existing in
isolated quantum wires into various phases of $2D$ quantum liquid. The 
most
drastic transformation is caused by the {\it interwire} tunneling 
$t_{\perp}$
in arrays of quantum wires with {\it intrawire} Coulomb repulsion.
This coupling constant rescales towards higher values for strong 
interaction
($g<1/2$), and the electrons in array transform into $2D$ Fermi liquid
\cite{Wen}.
The reason for this instability is the orthogonality catastrophe, i.e. 
the
infrared divergence in the low-energy excitation spectrum that 
accompanies
the interwire hopping processes.\\

Unlike interwire tunneling, the density-density or current-current
interwire interactions do not modify the low-energy behavior of quantum 
arrays
under certain conditions. In particular, it was shown recently
\cite{Luba00,Vica01,Luba01}
that an interaction of the type $W(n-n')$, which depends on the
distance between the wires $n$ and $n'$ but does not contain current
coordinates $x,x',$ imparts the properties of a {\it sliding phase} to 2D
array of 1D quantum wires. In this state an additional interwire coupling
leaves the  fixed-point action invariant under the "sliding" 
transformation
$\theta_n\to \theta_n+\alpha_n$ and $\pi_n \to \pi_n+\alpha^\prime_n$.
The contribution of interwire coupling reduces to a renormalization of 
the
parameters $v\to v(q_\perp)$, $g\to g(q_\perp)$ in the LL Hamiltonian
(\ref{D1}), where $q_\perp$ is a momentum perpendicular to the chain
orientation. Such LL structure can be interpreted as a quantum
analog of classical sliding phases of coupled $XY$
chains\cite{Hern}. Recently, it was found \cite{Sond} that a hierarchy of
quantum Hall states emerges in sliding phases when a quantizing magnetic 
field is applied to an array. \\  

In the present paper we concentrate on another aspect of the problem
of interacting quantum wires.  Instead of studying the conditions
under which the LL behavior is preserved in spite of interwire
interaction, we consider situations where the {\it dimensional
crossover} from $1D$ to $2D$ occurs.  Dimensional crossover is quite
well studied e.g. in thin semiconducting or superconducting films
where the film thickness is the control parameter that rules the
crossover (see e.g. Ref.\onlinecite {Buch}.  It occurs in strongly
anisotropic systems like quasi-one-dimensional organic
conductors\cite{sault} or layered metals\cite{metals}.  In the latter
cases temperature serves as a control parameter and crossover
manifests itself in interlayer transport.  In metals the layers appear
``isolated'' at high temperature, but become connected at low
temperatures to manifest $3D$ conducting properties.  Here we intend
to study another type of dimensional crossover, i.e. a {\it
geometrical} crossover, where the phase variable serves as a control
parameter, and the excitations in quantum array demonstrate either
$1D$ or $2D$ behavior in different parts of reciprocal space.\\

The most promising type of artificial structures where this effect is 
expected is a periodic $2D$ system of $m$ crossing arrays of parallel 
quantum wires.  We call it "quantum crossbars" (QCB).  The square 
grids of this type consisting of $2$ arrays were considered in various 
physical context in early papers \cite{Avr,Avi,Guinea,Castro,Kuzm}.  
In Refs.\onlinecite {Guinea,Castro} the fragility of the LL state against 
interwire tunneling in the crossing areas of QCB was studied.  It was 
found that a new periodicity imposed by the interwire hopping term 
results in the appearance of a low-energy cutoff $\Delta_l\sim \hbar 
v/a$ where $a$ is a period of the quantum grid.  Below this energy, 
the system is "frozen" in its lowest one-electron state.  As a result, 
the LL state remains robust against orthogonality catastrophe, and the 
Fermi surface conserves its 1D character in the corresponding parts of 
the $2D$ Brilllouin zone (BZ).  This cutoff energy tends to zero at 
the points where the one-electron energies for two perpendicular 
arrays $\epsilon_{k_{1}}$ and $\epsilon_{k_{2}}$ become degenerate.  
As a result, a dimensional crossover from $1D$ to $2D$ Fermi surface 
(or from LL to FL behavior) arises around the points 
$\epsilon_{F_{1}}=\epsilon_{F_{2}}$.\\

We study this dimensional crossover for Bose excitations (plasmons)
described by canonical variables $\theta,\pi$ in QCB. In order to
unravel the pertinent physics we consider a grid with {\it short-range
capacitive inter-wire interaction}.  This approximation seems natural
for $2D$ grids of carbon nanotubes \cite{Rueckes}, or artificially
fabricated bars of quantum wires with grid periods which exceed the
lattice spacing of a single wire or the diameter of a nanotube.  It
will be shown below that this interaction can be made effectively
weak.  Therefore, QCB retains the $1D$ LL character for motion along
the wires similarly to the case considered in Ref.\onlinecite{Luba01}.  At
the same time, the boson mode propagation along some resonant
directions is also feasible.  This is essentially a $2D$ process in
the $2D$ BZ (or in the elementary cell of the reciprocal
lattice).  \\

We start the studies of QCB with a double QCB $m=2$ (section 
\ref{sec:Double}).  In the first two subsections \ref{subsec:Notions} 
and \ref{subsec:Hamilt} we introduce basic notions and construct the 
Hamiltonian of the QCB. The main approximations are discussed in 
subsection \ref{subsec:Approx}.  Here we substantiate the used method 
(separable interaction approximation) and show that interaction 
between arrays in QCB is weak.  The energy spectra for square QCB and 
tilted QCB are described in detail in two parts 
\ref{subsubsec:Square} and \ref{subsubsec:Tilted} of subsection 
\ref{subsec:Spectr}.  Various correlation functions and related 
experimentally observable quantities (optical absorption, space 
correlators) are discussed in the last subsection \ref{subsec:Correl}.  
We predict here effect of peculiar ``Rabi oscillations'' - periodic 
energy transfer from one of the QCB array to another.\\

Triple QCB ($m=3$) formed by three arrays lying in parallel planes are
studied in Section \ref{sec:Triple}.  Such hexagonal grids may be
useful for three-terminal nanoelectronic devices \cite{Luo}.  The
plasmon spectra of triple QCB possess some specific features in
comparison with double QCB. We introduce the main notions and
construct the Hamiltonian of symmetric triple QCB (subsection
\ref{subsec:NotHam}), analyse the peculiarities of the frequency
spectrum (subsection \ref{subsec:SpecTriple}), and illustrate them by
description of triple Rabi oscillations - periodic energy transfer
between all three arrays (subsection \ref{subsec:Observ}).  The
results are summarized in Conclusion.  All technical details are
placed in Appendices A - E.\\

\section{Double QCB} \label{sec:Double}
\subsection{Basic notions}\label{subsec:Notions}
\begin{figure}[htb]
\centering
\includegraphics[width=75mm,height=40mm,angle=0,]{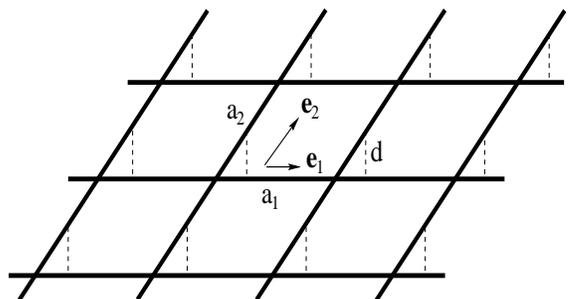}
\epsfxsize=70mm \caption{$2D$ crossbars formed by two
interacting arrays of parallel quantum wires. Here ${\bf e}_{1}, {\bf
e}_{2}$ are the unit vectors of the superlattice, ${a_{1}},{a_{2}}$ 
are the superlattice periods and $d$ is the vertical interarray 
distance}
\label{Bar3}
\end{figure}

Double QCB is a $2D$ periodic grid, which is formed by two 
periodically crossed arrays of $1D$ quantum wires.  In experimentally 
realizable setups \cite{Rueckes} these are cross-structures of 
suspended single-wall carbon nanotubes placed in two parallel planes 
separated by an inter-plane distance $d$.  However, some generic 
properties of QCB may be described in assumption that QCB is a genuine 
$2D$ system.  We assume that all wires of $j$-th array, $j=1,2,$ are 
identical.  They have the same length $L_j,$ Fermi velocity $v_{j}$ 
and Luttinger parameter $g_{j}.$ The arrays are oriented along the 
unit vectors ${\bf e}_{1,2}$ with an angle $\varphi$ between them.  
The periods of a crossbars along these directions are $a_{1}$ and 
$a_{2},$ and corresponding basic vectors are ${\bf a}_j=a_j{\bf e}_j$ 
(Fig.\ref{Bar3}).\\
The interaction between the
excitations in different wires is assumed to be concentrated around the
crossing points with coordinates $n_1{\bf a}_1+n_2{\bf
a}_2\equiv(n_1a_1,n_2a_2)$. The integers $n_j$ enumerate the wires
within the $j$-th array.  Such interaction imposes a superperiodicity
on the energy spectrum of initially one dimensional quantum wires, and
the eigenstates of this superlattice are characterized by a $2D$
quasimomentum ${\bf q}=q_1{\bf g}_1+q_2{\bf g}_2 \equiv(q_1,q_2)$. 
Here ${\bf g}_{1,2}$ are the unit vectors of the reciprocal
superlattice satisfying the standard orthogonality relations $({\bf
e}_i\cdot {\bf g}_j)=\delta_{ij}$.  The corresponding basic vectors of
the reciprocal superlattice have the form $m_1Q_1{\bf g}_1 +
m_2Q_2{\bf g}_2 $, where $Q_j=2\pi/a_j$ and $m_{1,2}$ are integers. \\

However the crossbars kinematics differs from that of a standard 2D periodic 
system. In conventional $2D$ systems, forbidden states in the inverse 
space arise due to Bragg diffraction in a $2D$ periodic potential, 
whereas the whole plane is allowed for wave propagation in real space, 
at least until the periodic potential is weak enough. A 
Brillouin zone is bounded by the Bragg lines. It coincides with a 
Wigner-Seitz cell of reciprocal lattice.  In sharply anisotropic QCB 
most of the real space is forbidden for electron and plasmon 
propagation.  The Bragg conditions for the wave vectors 
are modulated by a periodic potential  unlike  
those in conventional $2D$ plane.
These conditions are essentially one-dimensional.  Corresponding BZ is not a 
Wigner-Seitz cell of a reciprocal lattice but the elementary cell 
containing a site in its center.\\

\begin{figure}[htb]
\centering
\includegraphics[width=75mm,height=70mm,angle=0,]{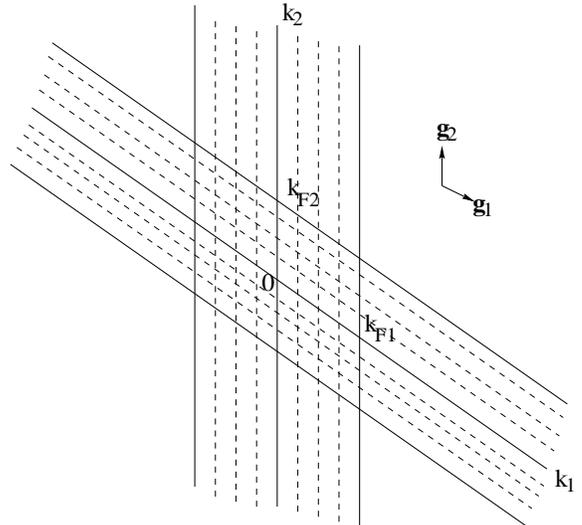}
\epsfxsize=70mm \caption{Fermi surface of $2D$ metallic quantum
bars in the absence of charge transfer between wires.  ${\bf
g}_{1}, {\bf g}_{2}$ are the unit vectors of the reciprocal
superlattice} \label{FS3}
\end{figure}

Indeed, the excitation motion in QCB is one-dimensional in major part 
of the $2D$ plane.  The anisotropy in real space imposes restrictions 
on the possible values of $2D$ coordinates $x_{1},x_{2}$ (${\bf 
r}=x_{1}{\bf e}_{1}+x_{2}{\bf e}_{2}$).  At least one of them, e.g., 
$x_2$ ($x_{1}$) should be an integer multiple of the corresponding 
array period $a_2$ ($a_{1}$), so that the vector ${\bf r}=(x_1,n_2 
a_2)$ (${\bf r}=(n_1a_1,x_2)$) characterizes the point with the $1D$ 
coordinate $x_1$ ($x_2$) lying at the $n_2$-th ($n_1$-th) wire of the 
first (second) array.  As a result, one cannot resort to the standard 
basis of $2D$ plane waves when constructing the eigenstate with a 
given wave vector ${\bf k}$. Even in {\it non-interacting} arrays of 
quantum wires (empty superlattice) the $2D$ basis is formed as a 
superposition of two sets of $1D$ waves.  The first of them is a set 
of $1D$ excitations propagating along {\it each} wire of the first 
array characterized by a unit vector $k_1{\bf g}_1$ with a phase shift 
$a_2k_2$ between adjacent wires.  The second set is the similar 
manifold of excitations propagating along the wires of the second 
array with the wave vector $k_2{\bf g}_2$ and the phase shift 
$a_1k_1$. The dispersion law of these excitations has the form
\begin{equation}
	\omega^{0}({\bf k})=\omega_{1}(k_{1})+\omega_{2}(k_{2}).\nonumber 
\end{equation}
The states of equal energy obtained by means of this 
procedure form straight lines in the $2D$ reciprocal space.  For 
example, the Fermi surface of QCB developed from the points $\pm 
k_{F1,2}$ for individual quantum wire consists of two sets of lines 
$|k_{1,2}|=k_{F1,2}$.  Respectively, the Fermi sea is not a circle 
with radius $k_F$ like in the case of free $2D$ gas, but a cross in 
the $k$ plane bounded by these four lines \cite{Guinea} (see 
Fig. \ref{FS3}). Finally, the Bragg conditions read
\begin{eqnarray*}
	\omega_{1}(k_{1}) & - & omega_{1}(k_{1}+ m_{1}Q_{1})\\
	& + & \omega_{2}(k_{2})-\omega_{2}(k_{2}+ m_{2}Q_{2})=0.
\end{eqnarray*}
and the lines $k_{1}=0$, $|k_{2}|=Q_{2}/2$, and $|k_{1}|=Q_{1}/2$, 
$k_{2}=0,$ satisfying these conditions, form a $2D$ BZ of double 
QCB.\\

Due to the inter-wire interaction, the excitations of QCB (see 
Figs.\ref{BZ2},\ref{BZ3} below) acquire genuine two-dimensionality 
characterized by the quasimomentum ${\bf q}=(q_1,q_2)$.  However, in 
case of weak interaction the $2D$ waves constructed from the $1D$ 
plane waves in accordance with the above procedure form an appropriate 
basis for the description of elementary excitations in QCB in close 
analogy with the nearly free electron approximation in conventional 
crystalline lattices. It is easily foreknown that a weak inter-wire 
interaction does not completely destroy the above quasimomentum 
classification of eigenstates, and the $2D$ reconstruction of the 
spectrum may be described in terms of wave mixing similarly to the 
standard Bragg diffraction in a weak periodic potential.  Moreover, 
the classification of eigenstates of empty superlattice may be 
effectively used for the classification of energy bands in a real QCB 
superlattice where the superperiodicity is imposed by interaction.\\

Complete kinematics of an empty superchain (wave functions, dispersion 
laws, relations between quasiparticle second quantization operators) 
is developed in Appendix A. In terms of these $1D$ Bloch functions 
(see Eqs.  (\ref{WaveFunc}), (\ref{WaveFunc1}) of Appendix A) we 
construct the $2D$ basis of Bloch functions for an empty superlattice
\begin{equation}
{\Psi}_{s,s',{\bf q}}({\bf r})={\psi}_{1,s,q_{1}}(x_{1})
            {\psi}_{2,s',q_{2}}(x_{2}).
            \label{Psi}
\end{equation}
Here $s,s'=1,2,\ldots,$ are the band numbers, and the
$2D$ quasimomentum ${\bf q}=(q_{1}, q_{2})$ belongs to the first BZ,
$|q_{j}|\leq Q_{j}/2.$ The corresponding eigenfrequencies are
\begin{equation}
    \omega_{ss'}({\bf q})=
    \omega_{1,s}({\bf q})+\omega_{2,s'}({\bf q}).\nonumber 
\end{equation}
Here
$$\omega_{j,s}({\bf q})\equiv \omega_{j,s}(q_{j}),$$
and $\omega_{j,s}(q_{j})$ is defined by Eqs. (\ref{Disp1}) below. 
We will use this basis in the next subsection when constructing the
excitation spectrum of QCB within the reduced band scheme.\\
\subsection{Hamiltonian}\label{subsec:Hamilt}
When turning to description of interaction in a QCB, one should refer 
to a real geometry of crossbars, and recollect the important fact that 
the equilibrium distance between two arrays is finite and large enough 
to supress direct electron tunneling 
cite{Rueckes}.  We neglect also the elastic and 
van der-Waals components of intertaction between real nanotubes, 
because these interactions are not involved in formulation of 
collective excitations in QCB. Then the full Hamiltonian of the QCB 
is
\begin{equation}
  H =  {H}_{1} + {H}_{2} +  H_{{int}},
  \label{TotHam1}
\end{equation}
where ${H}_{j}$ describes the $1D$ boson field
in the $j$-th array
\begin{eqnarray*}
{H}_{1} & = &
         \frac{\hbar{v_1}}{2}\sum_{{n}_{2}}
              \int\limits_{-L_1/2}^{L_1/2} {dx}_{1}
              \biggl\{
                   {g_1}{\pi}_{1}^{2} \left( x_1,
                   {n}_{2}a_2\right)
                   \\
               &+ &    \frac{1}{g_1}
                   \left(
                        {\partial}_{{x}_{1}}
                        {\theta}_{1}
                        \left(
                             {x}_{1},{n}_{2}a_2
                        \right)
                   \right)^2
         \biggr\},
\end{eqnarray*}
\begin{eqnarray*}
{H}_{2}& = &
     \frac{\hbar{v_2}}{2}\sum_{{n}_{1}}
         \int\limits_{-L_2/2}^{L_2/2}{dx}_{2}
         \biggl\{
          g_2{\pi}_{2}^{2}\left({n}_{1}a_1,{x}_{2}\right)
              \\
             &+ &\frac{1}{g_2}
              \left(
                   {\partial}_{x_2}
                   {\theta}_{2}
                   \left(
                        {n}_{1}a_1,{x}_{2}
                   \right)
              \right)^2
         \biggr\},
\end{eqnarray*}
and $\theta_j,\pi_j$
are the conventional canonically conjugated boson fields (see,
e.g., Ref.\onlinecite {Delft}).\\

\noindent
 The interwire interaction results from a short--range
contact capacitive coupling in the crosses of bars,
\begin{eqnarray*}
H_{{int}} & = & \sum\limits_{{n}_{1},{n}_{2}}
            \int dx_1
            dx_2
            V(x_1-n_1a_1,n_2a_2-x_2) \\
           & \times & {\rho}_{1}({x}_{1},n_2a_2)
            {\rho}_{2}(n_1a_1,{x}_{2}),
\end{eqnarray*}
where the integration is restricted by the area $-L_j/2\leq
x_j\leq L_j/2$. Here $\rho_i({\bf r})$ are density operators, and
$V({\bf r}_{1}-{\bf r}_{2})$ is a short-range interwire
interaction. Physically, it represents the Coulomb interaction
between charge fluctuations 
\begin{equation}
	e\zeta\left(\frac{x_j-n_{j}a_{j}}{r_{j}}\right), \ \ 
	\zeta(\xi)=\zeta(-\xi), \ \ 
    \zeta(0)=1,
	\label{charge}
\end{equation}
around the crossing point $(n_{1}a_{1},n_{2}a_{2}).$
The size of the fluctuation on the wire of the $j$-th array, 
is determined by the screening radius $r_{j}$ within the wire.
One may neglect the inter-wire tunneling and restrict oneself by
the capacitive interaction only, provided the vertical distance
between the wires $d$ is substantially larger than the screening
radiuses $r_{j}.$ Therefore the interaction has the form,
\begin{eqnarray*}
    V({\bf r})=\frac{V_{0}}{2}
        \Phi\left
        (\frac{x_1}{r_{1}},\frac{x_2}{r_{2}}
        \right),
\end{eqnarray*}
where the function $\Phi(\xi_{1},\xi_{2})$ is
\begin{equation}
    \Phi(\xi_{1},\xi_{2})=
        \frac
    {
    \displaystyle{
    \zeta_1(\xi_{1})
    \zeta_2(\xi_{2})
    }}
    {\displaystyle{
    \sqrt{1+\frac{|{\bf r}_{12}|^2}{d^2}
    }}
    },
        \label{F}
\end{equation}
$$
    {\bf r}_{12}= r_{1}\xi_{1}{\bf e}_{1}-
    r_{2}\xi_{2}{\bf e}_{2}.
$$
It is seen from these equations that $\Phi(\xi_{1},\xi_{2})$
vanishes for $|\xi_{1,2}|\ge 1$ and is normalized by condition
${\Phi}(0,0)=1$. The effective coupling strength is
\begin{equation}
    V_0=\frac{2e^{2}}{d}. 
    \label{strength}
\end{equation}
In terms of boson field operators
${\theta}_{i}$, the interaction is written as
\begin{eqnarray*}
    H_{{int}} & = & V_{0}\sum\limits_{{n}_{1},{n}_{2}}
            \int dx_1 dx_2
            \Phi\left
        (\frac{x_1-n_1a_1}{r_{1}},\frac{n_2a_2-x_2}{r_{2}}
        \right) {}\\
        &\times & {} \partial_{x_1}\theta_1(x_1,n_2a_2)
         \partial_{x_2}\theta_2(n_1a_1,x_2).
\end{eqnarray*}

In the quasimomentum representation
(\ref{Psi})
the full Hamiltonian (\ref{TotHam1}) acquires the form,
\begin{eqnarray}
H & = & \frac{{\hbar}{v}{g}}{2}\displaystyle{
                      \sum_{j=1}^{2}
                      \sum_{s,{\bf q}}
                      }
                      {\pi}_{js{\bf q}}^{\dagger}
                      {\pi}_{js{\bf q}}
                      +\nonumber\\
          & &\frac{\hbar}{2vg}\displaystyle{
          \sum_{jj'=1}^{2}
                 \sum_{s,s',{\bf q}}
                 }
                 {W}_{jsj's'{\bf q}}
                 {\theta}_{j s {\bf q}}^{\dagger}
                 {\theta}_{j's'{\bf q}},
                 \label{TotHam2}
\end{eqnarray}
where $\sqrt{vg/v_{j}g_{j}}{\theta}_{js{\bf q}}$ and 
$\sqrt{v_{j}g_{j}/vg}{\pi}_{js{\bf q}}$ are the Fourier components of 
the boson fields ${\theta}_{j}$ and ${\pi}_{j}$, and effective 
velocity and coupling are $v=\sqrt{v_1v_2}$, $g=\sqrt{g_1g_2}$ 
respectively.\\

The matrix elements for interwire coupling are given by:
\begin{eqnarray*}
{W}_{jsj's'{\bf{q}}} & = &
    {\omega}_{j s }(q_{j} )
    {\omega}_{j's'}(q_{j'})
    \left[
         {\delta}_{jj'}{\delta}_{ss'}+
         {\phi}_{jsj's'{\bf{q}}}
         \left(
              1-{\delta}_{jj'}
         \right)
    \right].
\end{eqnarray*}
Here
\begin{equation}
\omega_{js}(q_{j})=v_j
  \left(
       \left[\frac{s}{2}\right]Q_j+
       \left(-1\right)^{s-1}
       \left\vert{q}_{j}\right\vert
  \right),\nonumber
\end{equation}
are eigenfrequencies of the ``unperturbed'' $1D$ mode (see Eq.  
(\ref{Disp1}) of Appendix A), pertaining to an array $j$, band $s$ and 
quasimomentum ${\bf q}=q_{j}{\bf g}_{j}.$ The coefficients
\begin{eqnarray*}
{\phi}_{1s2s'{\bf{q}}} = \phi(-1)^{s+s'}
                       \mbox{sign}{(q_{1}q_{2})}
                       {\Phi}_{1s2s'{\bf{q}}},\\
                       \phi = 
                       \frac{gV_{0}r_{0}^{2}}{{\hbar}va},\ \ \
		       r_0=\sqrt{r_{1}r_{2}}, \ \ \
		       a=\sqrt{a_1a_2},
\end{eqnarray*}
are proportional to the dimensionless Fourier component of the
interaction strengths
\begin{eqnarray*}
{\Phi}_{1s2s'{\bf{q}}}& =&
  \int d{\xi}_{1}d{\xi}_{2}{\Phi}({\xi}_{1},{\xi}_{2})
  e^{-i(r_{1}q_1\xi_1+r_{2}q_2\xi_2)}\\
  & \times &
  u_{1,s, q_1}^{*}(r_{1}\xi_1)
  u_{2,s',q_2}^{*}(r_{2}\xi_2)={\Phi}^{*}_{2s'1s
  {\bf{q}}}.
\end{eqnarray*}

The Hamiltonian (\ref{TotHam2}) describes a system of coupled
harmonic oscillators, which can be {\em exactly} diagonalized with
the help of a certain canonical linear transformation (note that
it is already diagonal with respect to the quasimomentum ${\bf
q}$).  The diagonalization procedure is, nevertheless, rather
cumbersome due to the mixing of states belonging to different
bands and arrays. However, it will be shown below that provided
$d\gg{r}_{1,2}$, a separable potential
approximation is applicable, that shortens calculations noticeably.\\
\subsection{Approximations}\label{subsec:Approx}
As it was already mentioned, we consider the rarefied
QCB with short range capacitive interaction.  In the case of QCB
formed by nanotubes, this is a Coulomb interaction screened at a
distance of the order of the nanotube radius\cite{Sasaki} $R_{0},$
therefore $r_{0}\sim R_{0}.$ The minimal radius of a single-wall
carbon nanotube is about $R_{0}=0.35\div 0.4 nm$ (see
Ref.\onlinecite{Louie}).  The intertube vertical distance $d$ in
artificially produced nanotube networks is estimated as $d\approx
2$nm (see Ref.\onlinecite{Rueckes})  Therefore the ratio
$r_{0}^{2}/d^{2}\approx{0.04}$ is really small and {\it the
dimensionless interaction} $\Phi(\xi_{1},\xi_{2})$ (\ref{F}) {\it in the 
main approximation is separable}
\begin{equation}
  \Phi(\xi_{1},\xi_{2})\approx\Phi_{0}(\xi_{1},\xi_{2})= 
  \zeta_1(\xi_1)\zeta_2(\xi_2).
  \label{separ}
\end{equation}
It should be noted that the interaction in this form is
an even function of its arguments, and the odd correction to the
$\Phi_{0}$ is of order $r_0^2/d^2$, whereas $\Phi_{0}$ is of order of 
$1$.\\

To diagonalize the Hamiltonian (\ref{TotHam2}), one should solve the 
system of equations of motion for the field operators.  Generalized 
coordinates $\theta$ satisfy the equations
\begin{eqnarray}
  \left[\omega_{1s}^{2}(q_1)-\omega^{2}\right]
  \theta_{1s{\bf{q}}}+
  \sqrt{\varepsilon}\phi_{1s}(q_1)\omega_{1s}(q_1)
  \nonumber\\
  \times\frac{r_0}{a}\sum\limits_{s'}
  \phi_{2s'}(q_2)\omega_{2s'}(q_2)
  \theta_{2s'{\bf{q}}}=0,\nonumber\\
  s=1,2,\ldots ,
  \label{Euler-Lagr}
\end{eqnarray}
and the similar equations obtained by permutation 
$1\leftrightarrow 2$.  Here
\begin{equation}
  \phi_{js}(q)=(-1)^s\mbox{sign}(q)\int d\xi
  \zeta_{j}(\xi) e^{ir_{0}q\xi}u_{jsq}(r_{0}\xi),
  \label{FC-2}
\end{equation}
Bloch amplitudes $u_{jsq}(r_{0}\xi)$ are defined by Eqs. 
(\ref{WaveFunc1}) of Appendix A, and
\begin{equation}
    \varepsilon=\left(\phi\frac{a}{r_0}\right)^{2}=
    \left(\frac{gV_0r_0}{{\hbar}v}\right)^{2}.
    \label{epsilon}
\end{equation}
Due to separability of the interaction, equations of motion
(\ref{Euler-Lagr}) can be solved exactly.  Corresponding 
square eigenfrequencies are determined by the characteristic equation
\begin{equation}
 F_{1q_1}(\omega^2)F_{2q_2}(\omega^2)=\frac{1}{\varepsilon},
 \label{secul-eq}
\end{equation}
where
\begin{equation}
 F_{jq}(\omega^2)=\frac{r_{j}}{a_{j}}\sum\limits_{s}
 \frac{\phi_{js}^{2}(q)\omega_{js}^{2}(q)}
      {\omega_{js}^{2}(q)-\omega^2}.
 \label{F_j}
\end{equation}
The function $F_{jq}(\omega^2)$ has a set of poles at 
$\omega^2=\omega_{js}^{2}(q)$, $s=1,2,3,\ldots$ .  For squared 
frequency smaller than all squared initial eigenfrequencies 
$\omega_{js}^{2}(q)$, i.e.  within the interval $[0,\omega_{j1}^{2}]$, 
this is a positive and growing function.  Its minimal value $F_{j}$ on 
the interval is reached at $\omega^2=0,$ and it does not depend on 
quasimomentum $q$
\begin{equation}
F_{jq}(0)=\frac{r_{j}}{a_{j}}\sum\limits_{s}
 \phi_{js}^{2}(q)=\int d\xi \zeta_j^2(\xi)\equiv F_{j} 
\label{F_j0}
\end{equation}
(here Eqs.  (\ref{F_j}) and (\ref{FC-2}) are used). If parameter
$\varepsilon$ is smaller than its critical value
\begin{equation}
  \varepsilon_c=
  \frac{1}{F_{1}F_{2}},\nonumber 
\end{equation}
then all solutions $\omega^{2}$ of the characteristic equation are 
positive.  When $\varepsilon$ increases, the lowest QCB mode softens 
and its square frequency vanishes \textit{in a whole BZ} at 
$\varepsilon=\varepsilon_{c}$.  For exponential charge density 
distribution $\zeta(\xi)=\exp(-|\xi|),$ one obtaines 
$\varepsilon_c\approx 1$.\\

In our model the dimensionless interaction $\varepsilon$ in
Eq.(\ref{epsilon}) can be written as
\begin{equation}
    \varepsilon=\left(\frac{2R_{0}}{d}\frac{ge^{2}}{\hbar v}\right)^{2}.
    \label{epsilon1}
\end{equation}
For nanotube QCB, the first factor within parentheses is about 
$0.35.$ The second one which is nothing but the corresponding QCB 
``fine structure'' constant, can be estimated as $0.9$ (we used the 
values of $g=1/3$ and $v=8\times 10^{7}$cm/sec, see Ref.\onlinecite 
{Egger}).  Therefore $\varepsilon$ approximately equals $0.1,$ so this 
parameter is really small. Thus the considered system is stable, its 
spectrum is described by Eqs.(\ref{secul-eq}), (\ref{F_j}) with a 
\emph{small} parameter $\varepsilon$.\\

The general Eq.(\ref{secul-eq}) reduces in infrared limit ${\bf q}, 
\omega \to 0$ to an equation describing the spectrum of two coupled 
sliding phases.  i.e.  $1:1$ arrays in accordance with classification 
, offered in Ref.  \onlinecite{Luba01}.  Equation (3.13) of this paper 
is a long wave limit of our equation (\ref{omega-12}) derived in 
Appendix B. Therefore the general analysis of stability of the LL 
fixed point is appicable in our approach.  
\subsection{Spectrum}\label{subsec:Spectr}
Due to the smallness of interaction, the systematics of unperturbed 
levels and states is grossly conserved, at least in the low energy 
region corresponding to the first few energy bands.  This means that 
perturbed eigenstates could be described by the same quantum numbers 
(array number, band number and quasimomentum) as the unperturbed ones.  
Such a description fails in two specific regions of reciprocal space.  
The first of them is the vicinity of lines $q_{j}=nQ_{j}/2$ with $n$ 
integer.  Indeed, as it follows from the equations of motion 
(\ref{Euler-Lagr}), around these lines the interband mixing is 
significant.  These lines with $n=\pm 1$ include BZ boundaries.  
Because of this BZ which is, generally speaking, non relevant, and 
in this subsection we refer mostly to BZ.\\

The second region is the vicinity of the lines 
where the resonance conditions are fulfilled
\begin{equation}
	\omega^{2}_{1s}(q_{1})=\omega^{2}_{2s'}(q_{2}).
	\label{res}
\end{equation}
Here inter-array mixing within the same energy band ($s=s'$) or 
between neighboring bands ($s\neq s'$) is significant. In what 
follows we will pay attention first 
of all to these two regions because in the rest of the BZ the initial
systematics of the energy spectrum can be successfully used.\\

Equations (\ref{Euler-Lagr}), (\ref{secul-eq}), describing the wave 
functions 
 and the dispersion laws  are analysed in 
Appendix B. We describe below some of these dispersion 
qurves for two types of QCB basing on this analysis.\\
\subsubsection{Square QCB}\label{subsubsec:Square}
\begin{figure}[htb]
\centering
\includegraphics[width=75mm,height=70mm,angle=0,]{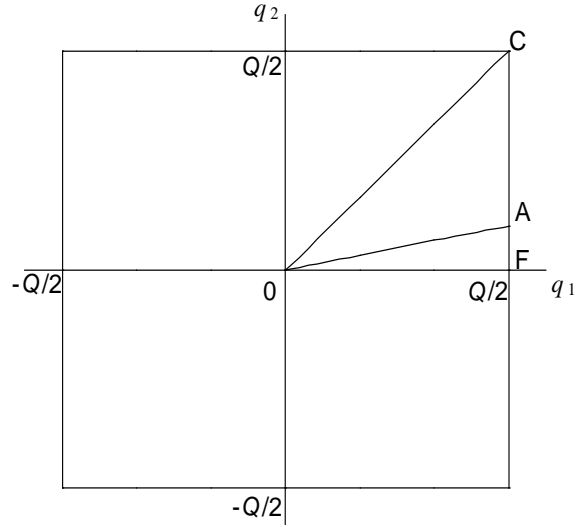}
\epsfxsize=70mm 
\caption{Two dimensional BZ of square QCB.} 
\label{BZ2}
\end{figure}

We start with the simplest case of square QCB formed by identical 
wires.  This means that all parameters (wire length, space period, 
Fermi velocity, LL parameter, screening radius) are the same for both 
arrays.  The corresponding BZ and is also a 
square (see Fig.\ref{BZ2}). Resonant lines are the diagonals of BZ.\\

In the major part of the BZ, for quasimomenta ${\bf{q}}$ lying far
from the diagonals, each eigenstate mostly conserves its initial
systematics, i.e. belongs to a given array, and mostly depends on a
given quasimomentum component.  Corresponding dispersion laws remain
linear being slightly modified near the BZ boundaries only.  The main
change is therefore the renormalization of the plasmon
velocity.\\
\begin{figure}[htb]
\centering
\includegraphics[width=75mm,height=55mm,angle=0,]{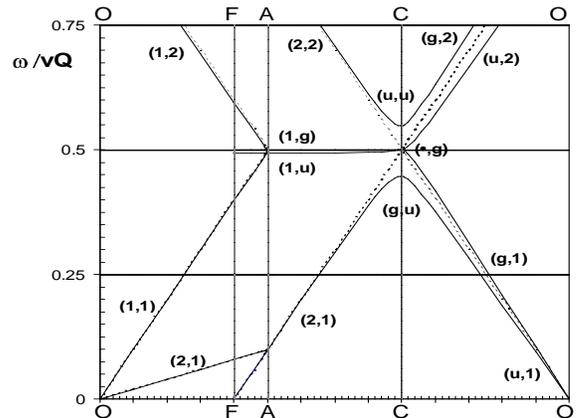}
\epsfxsize=70mm 
\caption{The energy spectrum of QCB (solid
lines) and noninteracting arrays (dashed lines) for quasimomenta at
the lines $OA,$ $FC,$ and $OC$ of BZ} 
\label{sp-sq}
\end{figure}

In the left part of Fig.\ref{sp-sq} we displayed dispersion curves
corresponding to quasimomenta belonging to a generic $OA$ line in
BZ.  In what follows we use $(j,s)$ notations for the
unperturbed boson propagating along the $j$-th array in the $s$-th
band.  Then the lowest curve in this part of Fig.\ref{sp-sq} is, in
fact, the slightly renormalized dispersion of a $(2,1)$ boson. The
middle curve describes $(1,1)$ boson, and the upper curve is the
dispersion of a $(1,2)$ boson.  The fourth frequency corresponding to
a $(2,2)$ boson, is far above and is not displayed in the figure.  It
is seen that the dispersion remains linear along the whole line $OA$
except a nearest vicinity of the BZ boundary (point $A$ in
Fig.\ref{BZ2}).\\

Dispersion curves corresponding to quasi momenta lying at the BZ 
boundary $q_{1}=Q/2,$ $0\leq q_{2}\leq Q/2$ (line $FC$ in 
Fig.\ref{BZ2}) are displayed in the central part in Fig.\ref{sp-sq}).  
The characteristic feature of this boundary is the intra-band 
degeneracy in one of two arrays.  Indeed, in zero approximation, two 
modes $(1,s),$ $s=1,2,$ propagating along the first array are 
degenerate with unperturbed frequency $\omega=0.5.$ The interaction 
lifts the degeneracy.  This interaction occurs to be repulsive at the 
BZ boundaries.  As a result the lowest of two middle curves in 
Fig.\ref{sp-sq} corresponds to $(1,u)$ boson, and upper of them 
describes $(1,g)$ boson.  Here the indices $g,u$ denote a boson parity 
with respect to the transposition of the band numbers.  Note that 
$(1,g)$ boson exactly conserves its unperturbed frequency 
$\omega=0.5.$ The latter fact is related to the square symmetry of the 
QCB.\\

Two others curves correspond to almost non pertubed bosons of the 
second array.  The lowest curve describes the dispersion of the 
$(2,1)$ wave.  Its counterpart in the second band $(2,2)$ is described 
by the highest curve in the figure.  Their dispersion laws are 
nearly linear, and deviations from linearity are observed only near 
the corner of the BZ (point $C$ in Fig.\ref{BZ2}).\\

Consider now dispersion relations of modes with quasi--momenta on the diagonal 
$OC$ of BZ and start with ${\bf q}$ not too close to the BZ corner 
$C$ ($q_{1}=q_{2}=Q/2$).  This diagonal is actually one of the 
resonance lines.  Two modes in the first band coressponding to 
different arrays are strongly mixed.  They mostly have a definite 
$j$-parity with respect to transposition of array numbers $j=1,2$.  
Interaction between these modes occurs to be attractive (repulsive) 
for $q_{1}q_{2}>0$ ($q_{1}q_{2}<0$).  Therefore the odd modes $(u,s),$ 
at the BZ diagonal $OC$ $s=1,2,$ correspond to lower frequencies and 
the even modes $(g,s)$ correspond to higher ones. The corresponding 
dispersion curves are displayed in the right part of 
Fig.\ref{sp-sq}.\\

 At the BZ corner $q_{1}=q_{2}=Q/2$ (point $C$ in Fig.\ref{BZ2}) all 
 four initial modes $j,s=1,2$ are degenerate in the lowest  
 approximation.  This four-fold degeneracy results from the square 
 symmetry of BZ (the resonant lines are diagonals of the Z).  Weak 
 inter-wire interaction partially lifts the degeneracy, however the 
 split modes have a definite $s$-parity with respect to transposition 
 of band numbers $s=1,2$.  The lowest frequency corresponds mostly to 
 $(g,u)$ boson, symmetric with respect to transposition of array 
 numbers, but antisymmetric with respect to the transposition of band 
 numbers.  The upper curve describes a $(u,u)$ boson with odd both 
 $j$-parity and $s$-parity.  The two middle modes with even band 
 parity, $(g,g)$ and $(u,g)$ bosons, remain degenerate and their 
 frequencies conserve the unperturbed value $\omega=0.5$.  This also 
 results from the square symmetry of QCB (\ref{F}).  \\
\begin{figure}[htb]
\centering
\includegraphics[width=65mm,height=65mm,angle=0,]{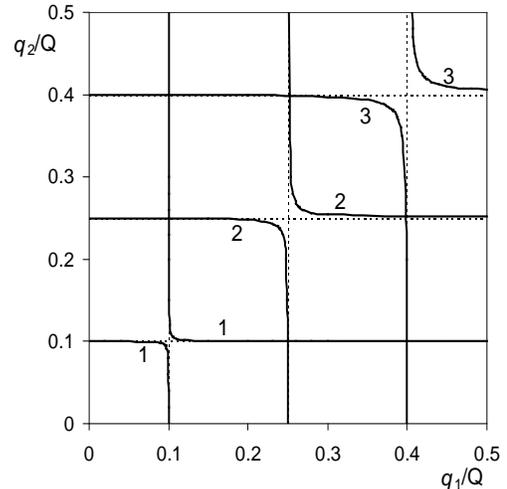}
\epsfxsize=80mm 
\caption{Lines of equal frequency of the lowest
mode for QCB (solid lines) and for noninteracting arrays (dashed
lines).  The lines $1,2,3$ correspond to the frequencies
$\omega_{1}=0.1,$ $\omega_{2}=0.25,$ $\omega_{3}=0.4$}
\label{IsoEn}
\end{figure}
All these results show that the quantum states of the $2D$ QCB
conserve the quasi $1D$ character of the Luttinger--like liquid in
major part of momentum space, and that $2D$ effects can be
successfully calculated within the framework of perturbation theory. 
However, bosons with quasimomenta close to the resonant line (diagonal
$OC$) of the BZ are strongly mixed bare $1D$ bosons.  These
excitations are essentially two-dimensional, and therefore the lines
of equal energy in this part of the BZ are modified by the $2D$
interaction (see Fig.\ref{IsoEn}).  It is clearly seen that deviations
from linearity occur only in a small part of the BZ. The crossover
from LL to FL behavior around isolated points of the BZ due to a
single-particle hybridization (tunneling) for Fermi excitations was
noticed in Refs.  \onlinecite{Guinea,Castro}, where a mesh of
horizontal and vertical stripes in superconducting cuprates was
studied.\\
\subsubsection{Tilted QCB}\label{subsubsec:Tilted}
\begin{figure}[htb]
\centering
\includegraphics[width=75mm,height=75mm,angle=0,]{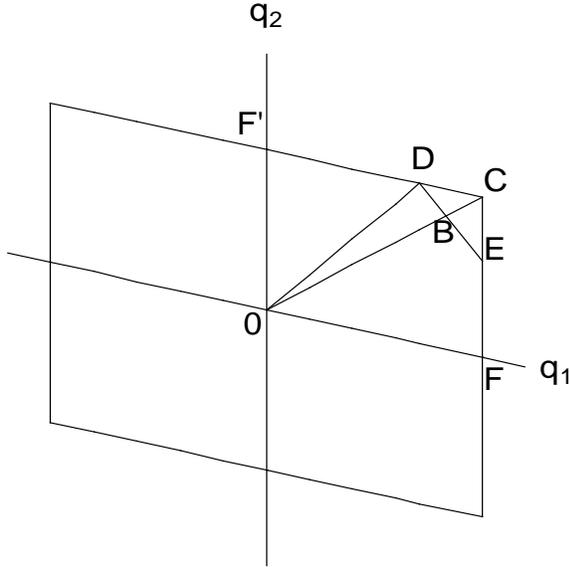}
\epsfxsize=70mm 
\caption{BZ of a titled QCB} 
\label{BZ3}
\end{figure}

Now we consider the spectrum of a generic double QCB. In this case all
parameters (wire length, space period, Fermi velocity, LL parameter,
screening radius) depend, generally speaking, on the array index $j.$
In what follows we refer to such a QCB as a tilted QCB. Now the
resonance condition (\ref{res}) is fulfilled not at the BZ diagonal
but at the resonant polygonal line.  Its part $ODE,$ lying in the
first quarter of the BZ, is displayed in Fig.\ref{BZ3} (all figures of
this subsection correspond to specific values
$v_{2}Q_{2}=1,\phantom{aa}v_{1}Q_{1}=1.4$).This results in
qualitative changes of the spectrum that are related first of all to
the appearance of two points $D$ and $E$ of the three-fold degeneracy for a
titled QCB (Fig.\ref{BZ3}) instead of a single point $C$ of four-fold
degeneracy for a square QCB (Fig.\ref{BZ2}).\\

\begin{figure}[htb]
\centering
\includegraphics[width=70mm,height=60mm,angle=0,]{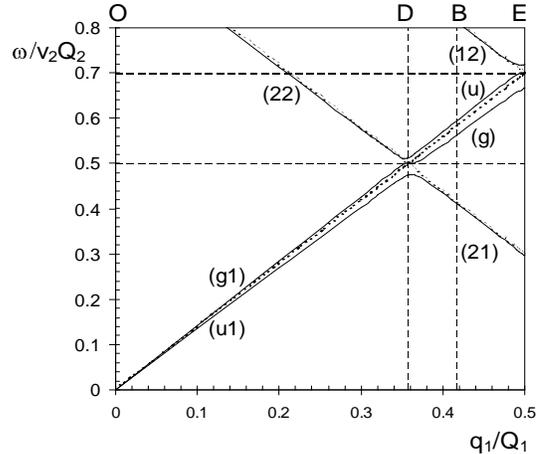}
\epsfxsize=80mm 
\caption{The energy spectrum of a tilted QCB (solid lines) and
noninteracting arrays (dashed lines) for quasimomenta on the resonant
line of the BZ (line $ODE$ in Fig.\ref{BZ3})}
\label{sp-res}
\end{figure}

We start with the resonant line $ODE$ (Fig.\ref{sp-res}).  The
dispersion curves at its $OD$ part and the symmetry properties of the 
corresponding eigenstates are similar to those at the $OC$ resonant line
for the square QCB (Fig.\ref{BZ2}).  The only difference is that instead
of the four-fold degeneracy at the BZ corner $C$ of the square QCB,
there is a three-fold degeneracy at the point $D$ lying at the BZ
boundary. A completely new situation takes place at the $DE$ line, where
two other modes $(1,1)$ and $(2,2),$ corresponding to different arrays
and different bands, are degenerate.  The interaction lifts this
degeneracy and the two middle lines in Fig.\ref{sp-res} describe even
$(g)$ and odd $(u)$ combinations of these modes.  The even mode
corresponds to the lowest frequency and the odd mode corresponds to the
higher one.  At the point $E$ one meets another type of a three-fold
degeneracy described in more detail in the next paragraph..\\

\begin{figure}[htb]
\centering
\includegraphics[width=70mm,height=60mm,angle=0]{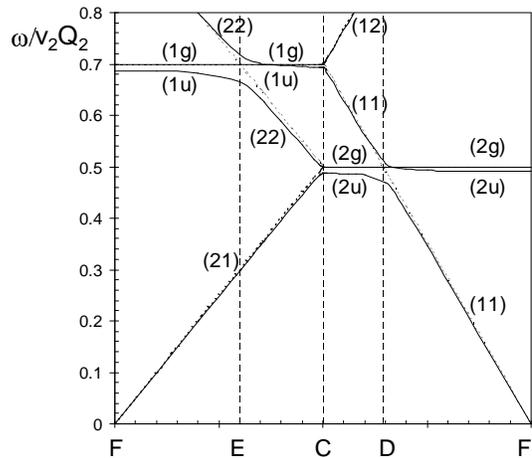}
\epsfxsize=80mm 
\caption{Energy spectrum of a tilted QCB (solid lines)
and noninteracting arrays (dashed lines) for quasimomenta at the
 BZ boundary (line $FCF'$ in the Fig.\ref{BZ3})}
\label{sp-bound}
\end{figure}

Dispersion curves corresponding to quasi momenta lying at the BZ
boundary $q_{1}=Q_1/2,$ $0\leq{q}_{2}\leq{Q}_{2}/2$ ($FC$ line in
Fig.\ref{BZ3}) and $q_{2}=Q_2/2,$ $0\leq{q}_{1}\leq{Q}_{1}/2$ ($CF'$
line in Fig.\ref{BZ3}), are displayed in Fig.\ref{sp-bound}.  The
lowest and the highest curves in the $FE$ part of the latter figure,
describe two waves propagating along the second array.  They are
nearly linear, and deviations from linearity are observed only near
the point $E$ where the interaction has a resonant character.  Two
modes propagating along the first array, in zero approximation, are
degenerate with an unperturbed frequency $\omega=0.7.$ The interaction
lifts the degeneracy.  The lowest of the two middle curves corresponds to
$(1,u)$ boson, and the upper of one describes $(1,g)$ boson.  Note that
$(1,g)$ boson conserves its unperturbed frequency
$\omega=0.7$.  The latter fact is related to the symmetry
$\zeta_j(\xi)=\zeta_j(-\xi)$ of the separable interaction (\ref{charge}). 
At the point $E$, the two modes propagating along the first array and
the mode propagating along the second array in the second band are
degenerate.  Interactions lifts the degeneracy, and, as a result,
the $(1,u)$ and $(2,2)$ waves are strongly mixed and the eigenmodes are
their even (highest frequency) and odd (lowest frequency)
combinations, and the $(1,g)$ mode (middle level).\\

There are two separate degeneracies within each array at the corner 
$C$ of a titled QCB BZ. Both of them are related to interband 
mixing conserving array index.  The spectral behavior along the $CF'$ 
boundary of the BZ is similar to that considered above but in the vicinity 
of the point $D$ of three-fold degeneracy.  Here, two modes 
propagating along the second array in the ßseparable potential 
approximation (\ref{separ}) remain degenerate.  This degeneracy is 
lifted only if deviation from separability is accounted for.\\
\begin{figure}[htb]
\centering
\includegraphics[width=70mm,height=65mm,angle=0,]{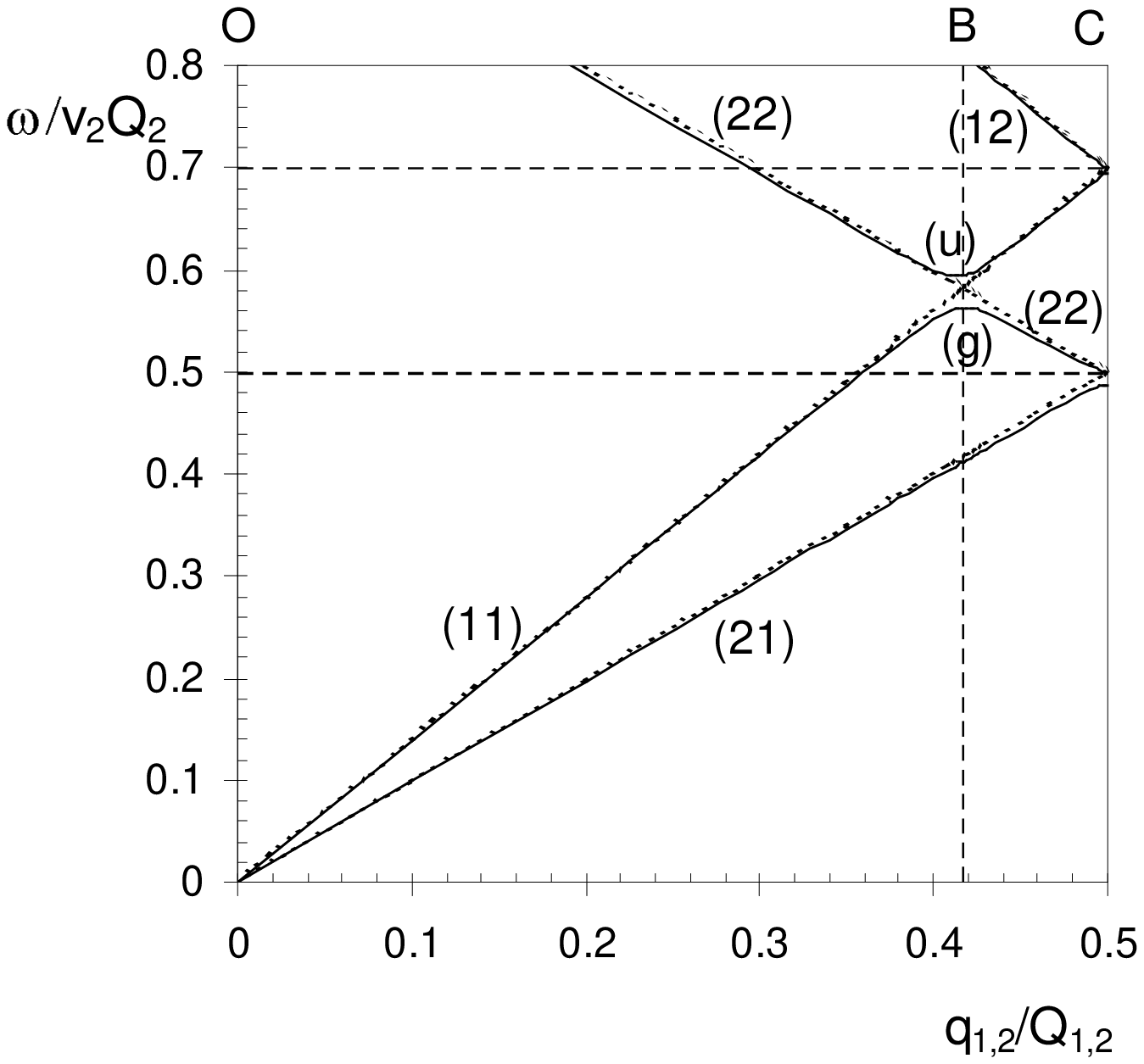}
\epsfxsize=80mm 
\caption{The energy spectrum of a titled QCB (solid lines) and
noninteracting arrays (dashed lines) for quasimomenta on the BZ diagonal
(line $OC$ in Fig.\ref{BZ3})}
\label{sp-diag}
\end{figure}
\begin{figure}[htb]
\centering
\includegraphics[width=85mm,height=75mm,angle=0,]{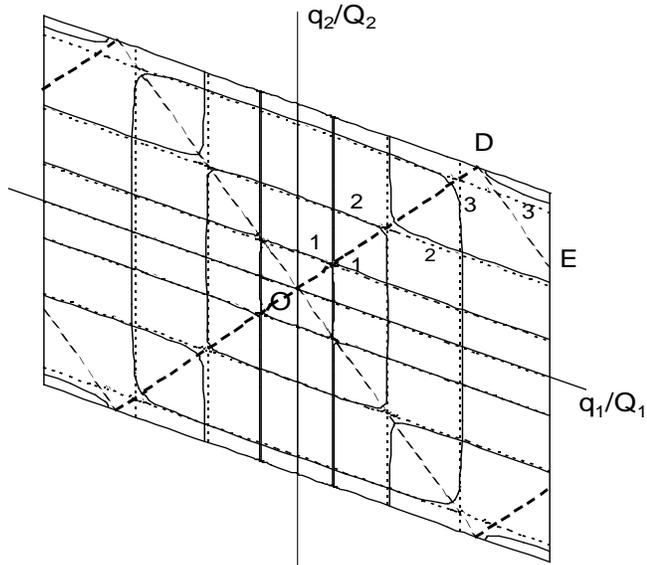}
\epsfxsize=80mm 
\caption{Lines of equal frequency for a tilted
QCB (solid lines) and noninteracting arrays (dashed lines). Lines
$1,2,3$ correspond to frequencies $\omega_{1}=0.1$,
$\omega_{2}=0.25$, $\omega_{3}=0.45$.}
\label{IsoEn1}
\end{figure}

The diagonal $OC$ of a tilted QCB BZ represents a new type of
generic line, that crosses a resonant line (Fig.\ref{sp-diag}).  Here
the spectrum mostly conserves its initial systematics, i.e. belongs to
a given array, and mostly depends on a given quasimomentum component. 
However, at the crossing point $B$, the modes $(1,1)$ and $(2,2),$
corresponding to both different arrays and bands, become degenerate
(two middle dashed lines in Fig.\ref{sp-diag}).  Interaction between
the wires lifts the degeneracy.  The eigenstates of QCB have a definite
parity with respect to transposition of these two modes.  The lowest
and upper of two middle lines corresponds to even ($g$) and odd ($u$)
mode, respectively.\\

\begin{figure}[htb]
\centering
\includegraphics[width=85mm,height=75mm,angle=0,]{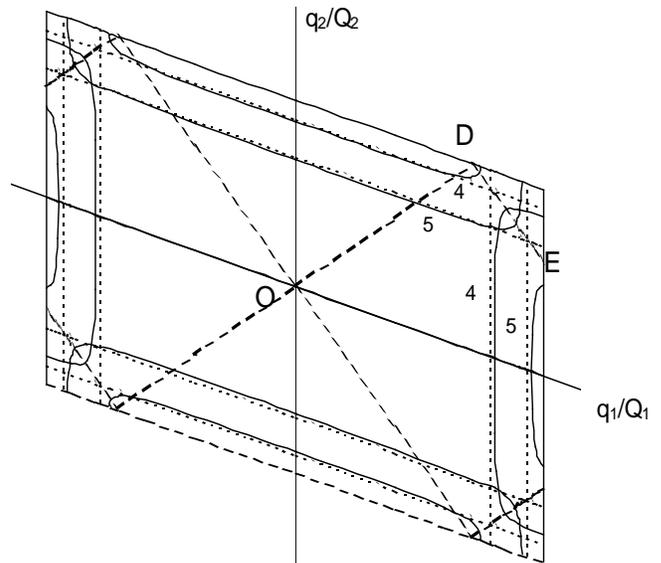}
\epsfxsize=80mm \caption{Lines of equal frequency for a tilted
QCB (solid lines) and noninteracting arrays (dashed lines). Lines 
$4,5$ in the lower panel correspond to frequencies
$\omega_{4}=0.55$, $\omega_{5}=0.65$}
\label{IsoEn2}
\end{figure}

Like in square QCB, bosons with quasimomenta close to the resonant 
lines are strongly mixed bare $1D$ bosons.  These excitations are 
essentially two-dimensional, and therefore lines of equal energy 
in the vicinity of the resonant lines are modified by the $2D$ 
interaction (see Figs.\ref{IsoEn1},\ref{IsoEn2}).  Deviations from 
$1D$ behaviour occur only in this small part of the BZ. For $\omega < 
0.5 v_2Q_2$ the lines of equal energy within BZ consist of closed line 
around the BZ center and four open lines (within the extendend bands 
scheme these lines are certainly closed) around the BZ corners (lines 
1, 2, 3 in Fig.\ref{IsoEn1}). At the line $OD$ in BZ, the modes of 
QCB are strongly coupled bare bosons propagating along both arrays in 
the first band.\\

For $0.5 v_2Q_2
<\omega < 0.5 v_1Q_1$ (lines 4, 5 in Fig.\ref{IsoEn2}) the topology of
lines of equal energy is modified.  In this case lines of
equal energy within the BZ consist of four open lines.  The
splitting of lines at the direction $DE$ corresponds to strong
coupling of modes propagating along the first array in the first band
with those propagating along the second array in the second band.\\ 

\subsection{Correlations and Observables}\label{subsec:Correl}
The structure of the energy spectrum analyzed above predetermines
optical and transport properties of QCB. We consider here three 
types of correlation functions manifesting dimensional crossover in 
QCB.\\
\subsubsection{Optical Absorption}\label{subsubsec:OptAbs}
We start with 
{\it ac} conductivity 
\begin{equation}
 {\sigma}_{jj'}({\bf{q}},\omega)=  
{\sigma}'_{jj'}({\bf{q}},\omega)+i{\sigma}''_{jj'}({\bf{q}},\omega).
\nonumber
\end{equation}
The real part ${\sigma}'_{jj'}({\bf{q}},\omega)$ determines an
optical absorption.  The spectral properties of \emph{ac} conductivity
are given by a current--current correlator
\begin{equation}
      {\sigma}_{jj'}({\bf{q}},\omega)=\frac{1}{\omega}
      \int\limits_{0}^{\infty}dt{e}^{i{\omega}t}
      \left\langle \left[
          {J}_{j1{\bf{q}}}(t),{J}_{j'1{\bf{q}}}^{\dag}(0)
      \right] \right\rangle.
	\label{CurrCorr}
\end{equation}
Here ${J}_{js{\bf{q}}}=\sqrt{2}vg{\pi}_{js{\bf{q}}}$ is a current
operator for the $j$-th array (we restrict ourselves
to the first band, for the sake of simplicity).\\

The current-current correlator for non-interacting wires is reduced to 
the conventional LL expression \cite{Voit},
\begin{eqnarray*}
\left\langle \left[
     {J}_{j1{\bf{q}}}(t),{J}_{j'1{\bf{q}}}^{\dag}(0)
\right] \right\rangle_0=
  -2ivg{\omega}_{j1{\bf{q}}}
  \sin({\omega}_{j1{\bf{q}}}t)
  {\delta}_{jj'}
\end{eqnarray*}
with metallic-like peak
\begin{equation}
{\sigma}'_{jj'}({\bf{q}},\omega>0)=
{\pi}vg
\delta({\omega}-{\omega}_{j1{\bf{q}}})
{\delta}_{jj'}.
\label{Drude_peak}
\end{equation}
For QCB this correlator is calculated in Appendix C. Its analysis 
leads to the following results.\\

The longitudinal absorption 
\begin{equation}
	\sigma'_{11}({\bf q},\omega)\propto 
	(1-\phi^{2}_{1{\bf q}})
	\delta(\omega-\tilde\omega_{1{\bf q}})+
	\phi^{2}_{1{\bf q}}
	\delta(\omega-\tilde\omega_{2{\bf q}})\nonumber
\end{equation}
contains well pronounced peak on the modified first array frequency 
and weak peak at the second array frequency (the parameter $\phi_{1{\bf 
q}},$ defined by Eq.  (\ref{phi1})of Appendix B, is small).  The 
modified frequencies $\tilde\omega_{1{\bf q}}$ and 
$\tilde\omega_{2{\bf q}}$ coincide with the eigenfrequencies 
$\omega_{+1{\bf q}}$ and $\omega_{-2{\bf q}}$ respectively, if 
$\omega_{1{\bf q}}>\omega_{2{\bf q}}.$ In the opposite case the signs 
$+,-$ should be changede to the opposite ones.\\

The transverse absorption 
component contains two weak peaks
\begin{equation}
	\sigma'_{12}({\bf q},\omega)\propto 
	\phi_{1{\bf q}}\left[
	\delta(\omega-\tilde\omega_{1{\bf q}})+
	\delta(\omega-\tilde\omega_{2{\bf q}})
	\right].\nonumber
\end{equation}

At the resonant line, the results change drastically. Both longitudinal 
and transverse components of the optical 
absorption contain two well pronounced peaks corresponding to 
slightly split modified frequencies
\begin{equation}
	\sigma'_{11}({\bf q},\omega)\propto\frac{1}{2} 
	\left[
	\delta(\omega-\tilde\omega_{1{\bf q}})+
	\delta(\omega-\tilde\omega_{2{\bf q}})
	\right].\nonumber
\end{equation}
\subsubsection{Space Perturbation}\label{subsubsec:Space}
One of the main effects specific for QCB is the appearance of
non-zero transverse momentum--momentum correlation function. In
space-time coordinates $({\bf{x}},t)$ its representation reads,
\begin{equation}
G_{12}({\bf x},t) =
           i\left\langle \left[
                             {\pi}_{1}(x_1,0;t),
                             {\pi}_{2}(0,x_2;0)
           \right] \right\rangle.\nonumber
\end{equation}
\begin{figure}[htb]
\centering
\includegraphics[width=75mm,height=75mm,angle=0,]{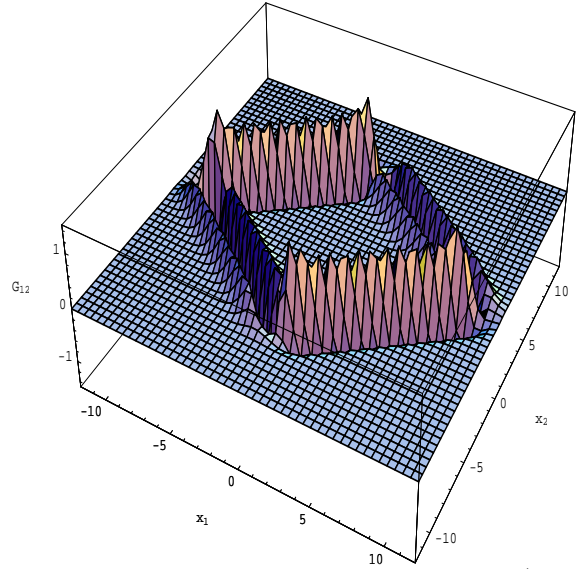}
\epsfxsize=80mm
\caption{The transverse correlation function $G_{12}(x_1,x_2;t)$
for $r_0=1$ and $vt=10$}
\label{GF12}
\end{figure}
This function describes the momentum response at the point $(0,x_{2})$
of the second array for time $t$ caused by an initial ($t=0$)
perturbation localized in coordinate space at the point $(x_{1},0)$
of the first array.  Standard calculations similar to those described
above, lead to the following expression,
\begin{eqnarray*}
& {G}_{12}({\bf{x}};t) =
                     \displaystyle{
                     \frac{V_0r_0^2}{4{\pi}^2{\hbar}}
                     \int\limits_{-\infty}^{\infty} dk_1 dk_2
                     }
                     {\phi}_1({k_1}){\phi}_2({k_2})k_1k_2
                      &
                     \\
                     & \times
                     \sin(k_1x_1)
                     \sin(k_2x_2)
                     \displaystyle{
                     \frac{v_2k_2\sin(v_2k_2t)-
                           v_1k_1\sin(v_1k_1t)}
                          {v_2^2k_2^2-v_1^2k_1^2},
                     } &
\end{eqnarray*}
where ${\phi}_{j}(k)$ is the form-factor (\ref{FC-2}) written in the
extended BZ. This correlator is shown in Fig.\ref{GF12}.  It is mostly
localized at the line determined by the obvious kinematic condition
$$\frac{|x_{1}|}{v_{1}}+\frac{|x_{2}|}{v_{2}}=t.$$
The time $t$ in the r.h.s.  is thea total time of plasmon propagation 
from the starting point $(x_{1},0)$ to the final point $(0,x_{2})$ or 
vice versa, along any of the shortest ways compatible with a 
restricted geometry of the $2D$ grid. The finiteness of the interaction 
radius slightly spreads this peak and modifies its profile.\\
\subsubsection{Rabi Oscillations}\label{subsubsec:Rabi}
Further manifestation of the 2D character of QCB system is related to 
the possibility of periodic energy transfer between the two arrays.  
Consider an initial perturbation which excites a plane wave with 
amplitude $\theta_{0}$ within the 
first array in the system of {\it non}-interacting arrays,
\begin{eqnarray*}
  {\theta}_{1}(x_1,n_2a_2;t)
  & = &
  \theta_{0}
  \sin(q_1x_1+q_2n_2a_2-v_1|q_1|t).
\end{eqnarray*}
If the wave vector
${\bf q},$ satisfying the condition $|{\bf{q}}|<<Q_{1,2}/2,$ is
not close to the resonant line of the BZ, weak interarray
interaction $\phi=\varepsilon{r}_{0}/{a}$ slightly changes the
$\theta_{1}$ component and leads to the appearance
of a small $\theta_{2}\sim\phi$ component.  But
for ${\bf q}$ lying on the resonant line
($v_1|q_1|=v_2|q_2|\equiv\omega_{\bf{q}}$), both components within
the main approximation have the same order of magnitude,
\begin{eqnarray*}
   {\theta}_{1}(x_1,n_2a_{2};t) & = &
   {\theta}_{0}
   \cos\left(
                 \frac{1}{2}
                 {\phi}_{1{\bf{q}}}
                 {\omega}_{{\bf{q}}}t
            \right)\\
   & \times &\sin(q_1x_1+q_2n_2a_2-{\omega}_{{\bf{q}}}t),
\end{eqnarray*}
\begin{eqnarray*}
   {\theta}_{2}(n_1a_1,x_2;t) & = &
   {\theta}_{0}
   \sin\left(
            \frac{1}{2}
            {\phi}_{1{\bf{q}}}
            {\omega}_{{\bf{q}}}t
   \right)\\
   & \times &
   \cos(q_1n_1a_1+q_2x_2-{\omega}_{{\bf{q}}}t).
\end{eqnarray*}
\begin{figure}[htb]
\centering
\includegraphics[width=70mm,height=60mm,angle=0,]{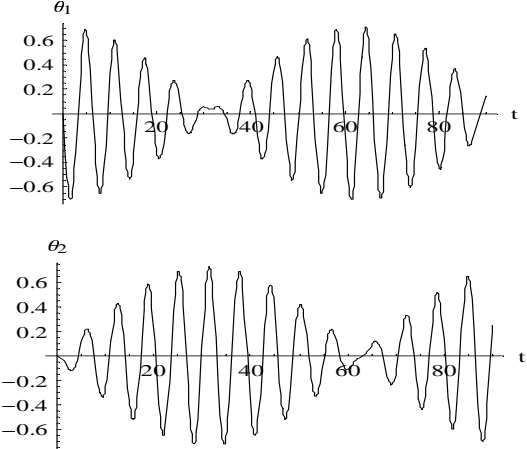}
\epsfxsize=80mm 
\caption{Periodic energy exchange between arrays (Rabi oscillations)}
\label{RO3}
\end{figure}
This corresponds to 2D propagation of a plane wave with wave vector 
${\bf q },$ {\it modulated} by a ``slow'' frequency $\sim\phi\omega.$ 
As a result, beating arises due to periodic energy transfer from one 
array to another during a long period $T\sim (\phi\omega)^{-1}$ (see 
Fig.\ref{RO3}).  These peculiar ``Rabi oscillations'' may be 
considered as one of the fingerprints of the physics exposed in QCB 
systems.\\
\section{Triple QCB}\label{sec:Triple}
\subsection{Notions and Hamiltonian}\label{subsec:NotHam}
\begin{figure}[htb]
\centering
\includegraphics[width=75mm,height=68mm,angle=0,]{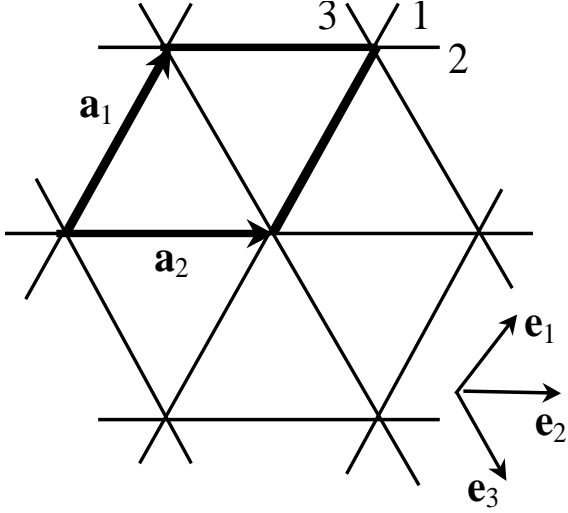}
\epsfxsize=70mm 
\caption{Triple QCB} 
\label{Bar4}
\end{figure}

Triple quantum bars is $2D$ periodic grid with $m=3$, formed by three 
periodically crossed arrays $j=1,2,3$ of $1D$ quantum wires.  In fact 
these arrays are placed on three planes parallel to $XY$ plane and 
separated by an inter-plane distances $d.$ The upper and the lower 
arrays correspond to $j=1,2,$ while the middle array has number $j=3.$ 
All wires in all arrays are identical.  They have the same length $L,$ 
Fermi velocity $v$ and Luttinger parameter $g.$ The arrays are 
oriented along the $2D$ unit vectors
\begin{equation}
	{\bf{e}}_{1}=\left(\frac{1}{2}, \frac{\sqrt{3}}{2}\right),
	\ \ \ \ {\bf{e}}_2=(1,0),\ \ \ \ 
	{\bf{e}}_{3}={\bf{e}}_2-{\bf{e}}_{1}.\nonumber
\end{equation}
The periods of QCB along these directions are equal, $a_{j}=a,$ so we 
deal with a regular triangular lattice.  In what follows we choose 
${\bf a}_{1,2}=a{\bf e}_{1,2}$ as the basic vectors of a superlattice 
(see Fig.\ref{Bar4}).\\

The wires within the $j$-th array are enumerated with the integers
$n_j.$ Define $2D$ coordinates along the $n_{j}$-th wire ${\bf r}_{j}$
as ${\bf r}_{j}=x_{j}{\bf{e}}_{j}+n_{j}a{\bf{e}}_{3}$ for upper and
lower arrays ($j=1,2$) and ${\bf
r}_{3}=x_{3}{\bf{e}}_{3}+n_{3}a{\bf{e}}_{1}$ for the middle array. 
Here $x_{j}$ are $1D$ continuous coorinates along the wire. The 
system of three non-interacting arrays is described by the Hamiltonian
\begin{equation}
	H_{0}=H_{1}+H_{2}+H_{3},\nonumber
\end{equation}
where
\begin{eqnarray}
 H_1 & = & \frac{{\hbar}v}{2}\sum\limits_{n_1}\int dx_1
 \left[
      g{\pi}_1^2(x_1{\bf{e}}_1+n_1a{\bf{e}}_3)\right.\nonumber\\
      &&+\left.\frac{1}{g}
      \left(
           \partial_{x_1}{\theta}_1(x_1{\bf{e}}_1+n_1a{\bf{e}}_3)
      \right)^2
 \right],
 \label{H1}
 \\
 H_2 & = & \frac{{\hbar}v}{2}\sum\limits_{n_2}\int dx_2
 \left[
      g{\pi}_2^2(x_2{\bf{e}}_2+n_2a{\bf{e}}_3)\right.\nonumber\\
      &&+\left.\frac{1}{g}
      \left(
           \partial_{x_2}{\theta}_2(x_2{\bf{e}}_2+n_2a{\bf{e}}_3)
      \right)^2
 \right],
 \label{H2}
 \\
 H_3 & = & \frac{{\hbar}v}{2}\sum\limits_{n_3}\int dx_3
 \left[
      g{\pi}_3^2(x_3{\bf{e}}_3+n_3a{\bf{e}}_1)\right.\nonumber\\
      &&+\left.\frac{1}{g}
      \left(
           \partial_{x_3}{\theta}_3(x_3{\bf{e}}_3+n_3a{\bf{e}}_1)
      \right)^2
 \right],
 \label{H3}
\end{eqnarray}
and $\pi_{j}$ and $\partial_{x_{j}}\theta_{j}$ are canonically
conjugated fields describing LL within the $j$-th array.\\
Interaction between the excitations in different wires of adjacent
arrays $j,j'$ is concentrated near the crossing points with
coordinates $n_j{\bf a}_j+n_{j'}{\bf a}_{j'}$.  It is actually Coulomb
interaction screened on a distance $r_0$ along each wire which is 
described by Hamiltonian
\begin{equation}
	H_{int}=H_{13}+H_{23},\nonumber
\end{equation}
where
\begin{eqnarray}
 &\displaystyle{\frac{H_{13}}{V_0}} =  \sum\limits_{n_1,n_3}\int dx_1dx_3
 \Phi\left(
          \frac{x_1-n_3a}{r_0}{\bf{e}}_1-
          \frac{x_3-n_1a}{r_0}{\bf{e}}_3
 \right)
 \nonumber\\  
 &\times
 \partial_{x_1}{\theta}_1(x_1{\bf{e}}_1+n_1a{\bf{e}}_3)
 \partial_{x_3}{\theta}_3(n_3a{\bf{e}}_1+x_3{\bf{e}}_3),
 \label{H13}
\end{eqnarray}
\begin{eqnarray}
 &\displaystyle{\frac{H_{23}}{V_0}}  = \sum\limits_{n_2,n_3}\int dx_2dx_3
\Phi\left(
          \frac{x_2-n_3a}{r_0}{\bf{e}}_2-
          \frac{x_3-n_2a}{r_0}{\bf{e}}_3
 \right)
 \nonumber\\  
 &\times
 \partial_{x_2}{\theta}_2(x_2{\bf{e}}_2+n_2a{\bf{e}}_3)
 \partial_{x_3}{\theta}_3(n_3a{\bf{e}}_2+x_3{\bf{e}}_3).
 \label{H23}
\end{eqnarray}
Here the effective coupling strength $V_{0}$ is defined by 
Eq.(\ref{strength}), the  
dimensionless interaction $\Phi$ is separable
\begin{equation}
 \Phi(\xi_{j}{\bf{e}}_{j}+\xi_{3}{\bf{e}}_{3})=
 \zeta(\xi_{j})\zeta(\xi_{3}),
 \ \ \ \ \ j=1,2,
 \label{separab}
\end{equation}
and $\zeta(\xi)$ is dimensionless charge fluctuation in the $j$-th 
wire (see Eq. (\ref{charge})).\\

\begin{figure}[htb]
\centering
\includegraphics[width=75mm,height=80mm,angle=0,]{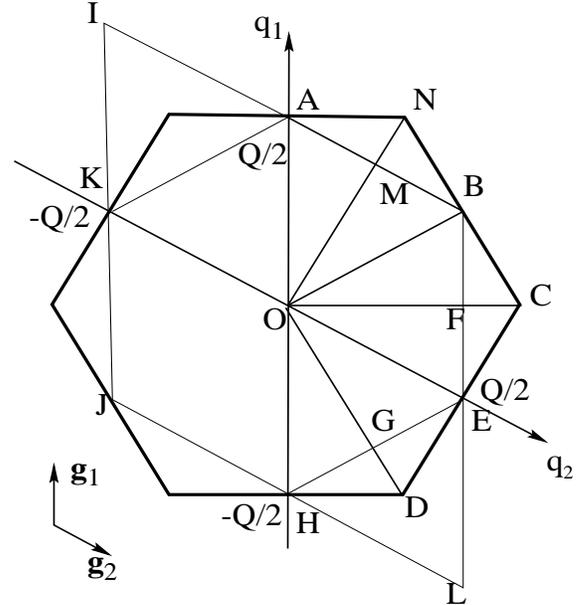}
\epsfxsize=70mm \caption{Elementary cell $BIJL$ of the reciprocal
lattice and the BZ hexagon of the triple QCB}
\label{BZ4}
\end{figure}

Such interaction imposes a superperiodicity on the energy spectrum of
initially one dimensional quantum wires, and the eigenstates of this
superlattice are characterized by a $2D$ quasimomentum ${\bf
q}=q_1{\bf g}_1+q_2{\bf g}_2 \equiv(q_1,q_2)$.  Here ${\bf g}_{1,2}$
are the unit vectors of the reciprocal superlattice satisfying the
standard orthogonality relations $({\bf e}_i\cdot {\bf
g}_j)=\delta_{ij}, \ \ j=1,2.$ The corresponding basic vectors of the
reciprocal superlattice have the form $Q(m_1{\bf g}_1 + m_2{\bf g}_2
$, where $Q=2\pi/a$ and $m_{1,2}$ are integers. In Fig.\ref{BZ4} 
elementary cell $BIJL$ of the reciprocal lattice is displayed together 
with the hexagon of the Wigner-Seitz cell that we choose as the  
BZ of the triple QCB.\\

To study the energy spectrum and the eigenstates of the total Hamiltonian
\begin{equation}
 	H=H_{0}+H_{int},
 	\label{H}
 \end{equation} 
we define the Fourier components of the field operators 
\begin{eqnarray}
 &{\theta}_1(x_1{\bf{e}}_1+n_1a{\bf{e}}_3) =\nonumber\\
 &(NL)^{-1/2}\sum\limits_{{s},{\bf{q}}}
 \theta_{1s{\bf{q}}}e^{i(q_1x_1+q_3n_1a)}
 u_{s,q_1}(x_1),
 \label{Fuorier1}
 \\
 &{\theta}_2(x_2{\bf{e}}_2+n_2a{\bf{e}}_3) =\nonumber\\
 &(NL)^{-1/2}\sum\limits_{{s},{\bf{q}}}
 \theta_{2s{\bf{q}}}e^{i(q_2x_2+q_3n_2a)}
 u_{s,q_2}(x_2),
 \label{Fuorier2}
 \\
 &{\theta}_3(x_3{\bf{e}}_3+n_3a{\bf{e}}_1) =\nonumber\\
 &(NL)^{-1/2}\sum\limits_{{s},{\bf{q}}}
 \theta_{3s{\bf{q}}}e^{i(q_3x_3+q_1n_3a)}
 u_{s,q_3}(x_3).
 \label{Fuorier3}
\end{eqnarray}
Here 
$${\bf q}=q_{1}{\bf e}_{1}+q_{2} {\bf e}_{2},\ \ \ 
q_{3}=q_{2}-q_{1},$$
and $N=L/a$ is the dimensionless length of a wire.  In the ${\bf q}$ 
representation, the Hamiltonians $H_{j}$ (Eqs. (\ref{H1})-(\ref{H3}))  
and $H_{j3}$ (Eqs. (\ref{H13}), (\ref{H23})) can be written as
\begin{eqnarray*}
 H_j  =  \frac{{\hbar}{v}{g}}{2}\sum\limits_{{s},{\bf{q}}}
           \pi_{js{\bf{q}}}^{+}\pi_{js{\bf{q}}}
		   \ \ \ \ \ \ \ \ \ \ \ \ \ \ \ \ \ \ \ \ \ \ \ \ \ \ \ 
		   \\
           +\frac{\hbar}{{2}{v}{g}}\sum\limits_{{s},{\bf{q}}}
           {\omega}_{s}^{2}(q_j)
           \theta_{js{\bf{q}}}^{+}\theta_{js{\bf{q}}},
 \ \ \ \ \ \ \ j=1,2,3,
 \\
 H_{j3}  =  \frac{V_0r_0^2}{{2}{v}{g}}\sum\limits_{s,s',{\bf{q}}}
 \phi_{s}(q_3)\phi_{s'}(q_j)\omega_{s}(q_3)\omega_{s'}(q_j)
 \\
 \times\left[
      \theta_{3s{\bf{q}}}^{+}\theta_{js'{\bf{q}}}+h.c.
 \right], \ \ \ \ j=1,2,
\end{eqnarray*}
where 
$$
\omega_s(q)=v\left(\left[\frac{s}{2}\right]Q+
\left(-1\right)^{s-1}|q|\right).
 \ \ \ \ \ Q=\frac{2{\pi}}{a}
$$
Thus the total Hamiltonian (\ref{H}) describes a system of coupled 
harmonic oscillators, and can be diagonalized exactly like in the case 
of double QCB.\\

\subsection{Spectrum}\label{subsec:SpecTriple}
Separability of the interaction (\ref{separab}) allows one to 
derive analytical equations for the spectrum of the total 
Hamiltonian (\ref{H}) (see Appendix D).  Here we 
describe the behavior of the spectrum and the states along some 
specific lines of the reciprocal space.\\

The high symmetry of the triple QCB leads to a number of lines where
interarray or interband resonant interaction occurs: \emph{all} lines
in Fig. \ref{BZ4} posess some resonant properties. These lines 
may be classified as follows:

On the Bragg lines where one of three array wavenumbers $q_{j}$ is a
multiple integer of $Q/2,$ there is a strong intraband mixing of modes
of the $j$-th array.  In Fig.\ref{BZ4}, these lines are the boundaries
of the elementary cell of the reciprocal lattice $IJLB,$ axes $q_{1}$
and $q_{2}$, lines $OB$ and $EH.$ In particular, along the lines
$OA$ ($q_{2}=0$) and $OB$ ($q_{3}=0$) two modes corresponding to $2$-d
and $3$-d bands and to the second ($OA$) or third ($OB$) array are
mixed.  Along the line $AB$ ($q_{1}=Q/2$) the same mixing happens
between $(1,1)$ and $(1,2)$ modes.  Moreover, the resonant mixing of
different arrays within the same band occurs along the medians $OA,$
$OB,$ etc..  There are two types of such a resonance.  The first of
one (e.g. $OA$ line) is the resonance between neighboring arrays
($q_{1}=-q_{3}$) and therefore it is of the main order with respect to
interaction.  The second one (e.g. $OB$ line) is the resonance between
remote arrays ($q_{1}=q_{2}$) and is one order smaller.\\

The second family consists of resonant lines formed by the BZ hexagon
boundaries and diagonals.  Thus, the diagonal $OC$ realizes a first
order resonance between the first and the third arrays $q_{1}=q_{3},$
and the BZ boundaries $HD$ and $AN$ correspond to the same resonance
up to an umklapp process ($q_{1}=q_{3}-Q$ and $q_{1}=q_{3}+Q$
respectively).  Along the diagonal $OD$ and the BZ boundary $NC$ a
second order resonance takes place with resonance conditions
$q_{2}=-q_{1}$ and $q_{2}=-q_{1}+Q$ respectively.\\
 
In the reciprocal space of the triple QCB there are four different
types of crossing points.  Two of them include the bases of BZ medians
(e.g. points $A,$ $B,$ $E$ and so on).  Here one deals with the
four-fold degeneracy of the modes corresponding to the first order
resonance between the neighboring arrays (e.g. point $A,$
$\omega_{1,s}=\omega_{3,s'},$ $s,s'=1,2$), or to the second order
resonance between remote arrays (like point $B,$
$\omega_{1,s}=\omega_{2,s'},$ $s,s'=1,2$).  One more family consists
of crossing points of the BZ diagonals and the lines connecting the
bases of its medians (points $M,$ $F,$ $G$ and so on).  Here one deals
with three types of two-fold degeneracy simultaneously.  For
example, at the point $M$ two separate pairs of modes corresponding to
neighboring arrays $(2,1),$ $(3,1),$ and $(2,2),$ $(3,2),$ are
degenerate, as well as two modes corresponding to the first array,
$(1,1),$ $(1,2).$ Finally the BZ hexagon vertices form the most
interesting group of points where the three-fold degeneracy between
modes corresponding to all three arrays takes place.  The typical
example of such a point is the vertex $C$ where the resonance condition
$q_{1}=-q_{2}+Q=q_{3}=Q/3$ is satisfied.\\
\begin{figure}[htb]
\centering
\includegraphics[width=75mm,height=60mm,angle=0,]{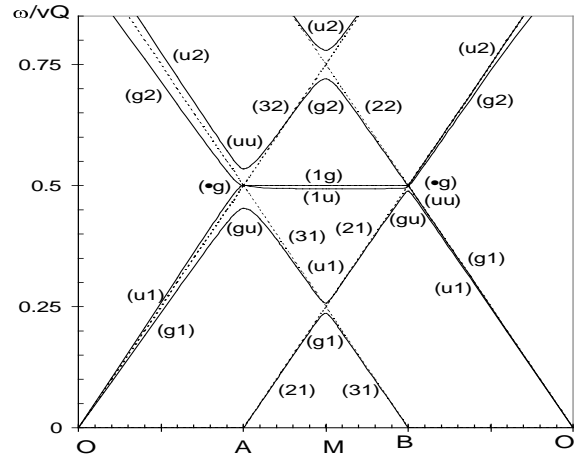}
\epsfxsize=70mm 
\caption{Dispersion curves at the $OAMBO$ polygon of BZ} 
\label{Tri1}
\end{figure}

Almost all these peculiarities of the triple QCB spectrum can be
illustrated in Fig.\ref{Tri1} where the dispersion curves along the
closed line $OABO$ are displayed.  We emphasize once more that in the
infrared limit $\omega,{\bf q}\to 0$ triple QCB like double QCB
preserves the characteristic LL properties of the initial arrays.\\
\subsection{Observables}\label{subsec:Observ}
The structure of the energy spectrum analyzed above strongly 
influences optical and transport properties of the triple QCB. As in 
the case of double QCB (subsection \ref{subsec:Correl}), one expects to 
observe four peaks of the optical absorption near the points $A,B,E,H$ 
of the four-fold degeneracy.  Then, specific features of space correlators 
like those considered in \ref{subsubsec:Space} can be observed.  But the 
most pronounced manifestation of a triangular symmetry of the triple 
QCB are its Rabi oscillations.\\

\begin{figure}[htb]
\centering
\includegraphics[width=75mm,height=105mm,angle=0,]{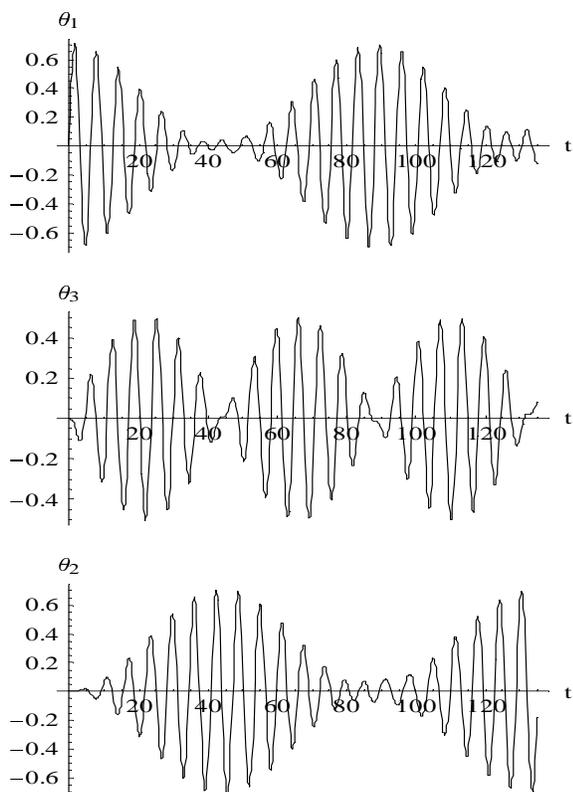}
\epsfxsize=70mm 
\caption{Periodic energy transfer between three arrays at the triple 
resonant point $C$ of the BZ} 
\label{ROTr}
\end{figure}

Consider the vicinity of the point $C$ of three-fold degeneracy mixing 
all three arrays.  Appropriate initial conditions lead (see Appendix E 
for details) to the following time dependence of the field operators 
in the coordinate origin in real space
\begin{eqnarray}
  \theta_1(0,0;t) &=& \theta_0
  \sin(\omega_0t)
  \cos^2\left(\frac{\sqrt{2\varepsilon}\phi^2}{4}\omega_0t\right),
  \nonumber\\
  \theta_2(0,0;t) &=&\theta_0
  \cos(\omega_0t)
  \sin^2\left(\frac{\sqrt{2\varepsilon}\phi^2}{4}\omega_0t\right),
  \nonumber\\
  \theta_3(0,0;t) &=& \theta_0
  \sin(\omega_0t)
  \cos\left(\frac{\sqrt{2\varepsilon}\phi^2}{2}\omega_0t\right).
  \label{sol3}
\end{eqnarray}
The field operators of all three arrays demonstrate fast oscillations
with the resonant frequency $\omega_{0}$ modulated by a slow
frequency.  It is the same for the two remote arrays, and doubled for
the intermediate array.  These beatings are synchronized in a sense
that zero intensity on the intermediate array always coincides with
the same intensity on one of the remote arrays.  At these moments all
the energy is concentrated solely within one of the remote arrays. 
These peculiar Rabi oscillations are displayed in Fig.\ref{ROTr}.\\

\section{Conclusion}\label{sec:Concl}
We discussed in this paper the kinematics and dynamics of plasmon
spectrum in QCB. These nanostructures may be fabricated from
single-wall carbon nanotubes \cite{Rueckes,Diehl}.  On the one hand,
QCB is promised to become an important component of future molecular
electronics \cite{Rueckes,Tseng}.  On the other hand, the spectrum of
elementary excitations (plasmons) in these grids possesses the
features of both 1D and 2D electron liquid.  As is shown in Refs. 
\onlinecite{Luba00,Luba01} and confirmed in the present study, the energy
spectrum of QCB preserves the characteristic properties of LL at
$|{\bf q}|,\omega\to 0,$.  At finite ${\bf q},\omega$ the density and
momentum waves in double and triple QCB may have either $1D$ or $2D$
character depending on the direction of the wave vector.  Due to
interarray interaction, unperturbed states, propagating along arrays
are always mixed, and transverse components of correlation functions
do not vanish.  For quasi-momentum lying on the resonant lines of the BZ,
such mixing is strong and transverse correlators have the same order
of magnitude as the longitudinal ones.  Periodic energy transfer
betweem arrays (``Rabi oscillations'') is predicted.\\

The crossover from 1D to 2D regime may be experimentally observed. 
One of the experimental manifestations, i.e. the crossover from
isotropic to anisotropic (spatially nonuniform) conductivity was
pointed out in Ref.  \onlinecite {Luba01}.  The current may be
inserted in QCB at a point on an array $j$ and extracted from another
array $i$ at a distance $r$.  Then a temperature dependent length
scale $l(T)$ arises, so that for $r\gg l$ the resistance is dominated
by small $q$ and therefore, the current is isotropic.  In the opposite
limit $r<l$ the dependence of the current on the points of
injection/extraction may be detected.  At $T=0$ the length $l$ becomes
infinite, and current can only be carried along the wires.  These
effects are in fact manifestations of the LL behavior of the QCB in
the infrared limit.\\

To observe the crossover at finite $\{\omega, {\bf q}\}$, one should
find a way of exciting the corresponding plasmon modes.  Then,
scanning the $\omega(q_1,q_2)$ surfaces, one may in principle detect
the crossover from quasi 1D to 2D behavior in accordance with the
properties of the energy spectra presented in Sections II and III.
Plasmons in QCB may be excited either by means of microwave resonators
or by means of interaction with surface plasmons.  In the latter case one
should prepare the grid on a corresponding semiconductor substrate and
measure, e.g., the plasmon loss spectra.  The theory of these plasmon
losses will be presented in a forthcoming publication.  \\
\section*{Acknowledgements}
This research is supported in part by grants from the Israel
Science foundations, the DIP German-Israel cooperation program, and the
USA-Israel BSF program.\\
\section*{Appendix A. Empty Superchain}\label{sec:Empty}
Here we construct eigenfunctions, spectrum, and quasi-particle operators 
for an ``empty superchain'' - quantum wire in an infinitely weak 
periodic potential with period $a$. Excitations in an initial wire are
described as plane waves $L^{-1/2}\exp (ikx)$ with wave number 
$k=2\pi n/L,$ with integer $n,$ and
dispersion law $\omega(k)=v|k|$ (the array number is
temporarily omitted). The following orthogonality relations are valid
\begin{eqnarray*}    
    \int_{-L/2}^{L/2}\psi_{k}^{*}(x)\psi_{k'}(x)dx & = & \delta_{k,k'},
    \\
	\sum_{k}\psi_{k}^{*}(x)\psi_{k}(x') & = & \delta_{L}(x-x'),
\end{eqnarray*}
where $\delta_{L}$ stands for  periodic delta-function
$$\delta_{L}(x-x') \equiv \sum_{n}\delta (x-x'-nL).$$ 

``Empty superchain'' is characterized by a space period $a$ and 
corresponding reciprocal lattice wave number $Q=2\pi/a.$ Each 
excitation in such a superchain is described by its quasi-wavenumber 
$q$ and a band number $s$ ($s=1,2,\ldots$) that are related to the 
corresponding wave number $k$ by the following relation,
\begin{equation}
	k=q+iQx(-1)^{s-1}\left[\frac{s}{2}\right]
    {\mbox{ sign }}q.\nonumber
\end{equation}
The corresponding wave 
function $\psi_{s,q}(x)$ has the Bloch-type structure,
\begin{equation}
    \psi_{s,q}(x)=\frac{1}{\sqrt{L}}
    e^{iqx}
    u_{s,q}(x),
    \label{WaveFunc}
\end{equation}
and satisfies the orthogonality relations
\begin{eqnarray*} 
	\int_{-L/2}^{L/2}\psi_{s,q}^{*}(x)\psi_{s',q'}(x)dx & = &\delta_{s,s'}
	\delta_{Q;q,q'}, 
	\\ 
	\sum_{s,q}\psi_{s,q}^{*}(x)\psi_{s,q}(x') & = & \delta_{L}(x-x'),
\end{eqnarray*}
where
$$\delta_{Q;q,q'}=\sum_{n}\delta_{q+nQ,q'}.$$
Within the first BZ, $-Q/2\leq q<Q/2,$ Bloch amplitude and dispersion 
law $\omega_{s}$ have the following form
\begin{eqnarray}
    u_{s,q}(x) & = &
    \exp
    \left\{
    iQx(-1)^{s-1}\left[\frac{s}{2}\right]
    {\mbox{ sign }}q
    \right \},
    \label{WaveFunc1}\\
    \omega_{s}(q) & = &  vQ\left(
    \left[\frac{s}{2}\right]
    +(-1)^{s-1}\frac{|q|}{Q}
    \right).
    \label{Disp1}
\end{eqnarray}
Here square brackets denote an integral part of a number. 
Taking into account that both Bloch amplitude $u_{s,q}(x)$ and 
dispersion law $\omega_{s}(q)$ are periodic functions of $q$ with 
period $Q,$ one obtaines general equations for the Bloch amplitude
\begin{eqnarray*}
	    u_{s,q}(x) & = &
    \displaystyle{
    \sum_{n=-\infty}^{\infty}
    \frac{\sin\xi_{n}}{\xi_{n}}}\\
    && \times\cos[(2s-1)\xi_{n}]
    \exp\left(-4i\xi_{n}\frac{q}{Q}\right),\\
    4\xi_{n} & = & Q(x-na).
\end{eqnarray*}
and dispersion law $\omega_{s}(q)$
\begin{eqnarray*}
	(vQ)^{-1}\omega_{s}(q) & = &
    \frac{2s-1}{4}\\
    & + &
    \displaystyle{\sum_{n=1}^{\infty}}
    \frac{2 (-1)^{s}}{\pi^{2}(2n+1)^{2}}
        \cos \frac{2\pi (2n+1)q}{Q}.
\end{eqnarray*}
The relations between quasiparticle operators for a free wire, 
$c_{k},$ for momentum $k\neq nQ/2$ with $n$ integer, and those for an 
empty superchain, $C_{s,q},$ for quasimomentum $q$ from the first BZ, 
$-Q/2<q<Q/2,$ look as
\begin{eqnarray*}
	c_{k}=C_{s,q}{\mbox{sign}}k,\\
	s=1+\left[\frac{2|k|}{Q}\right], \ \ \ 
	q=Q\left(\left\{\frac{k}{Q}
	+\frac{1}{2}\right\}-\frac{1}{2}\right)
\end{eqnarray*}
\begin{eqnarray*}
    C_{s,q}=(-1)^{\nu}c_{k},\\
	k=q+(-1)^{\nu}Q
	\left[\frac{s}{2}\right], \phantom{aaa} 
	\nu=s+1+\left[\frac{2q}{Q}\right],
\end{eqnarray*}
where curely brackets denote a fractional part of a number. For  
obtaining these relations we used the folowing expression
\begin{eqnarray*}
	\int_{-L/2}^{L/2}\psi_{k}^{*}(x)\psi_{s,q}(x)dx=
	\delta_{s,s(q)}\delta_{Q;q,k}{\mbox{sign}}k,\\
	s(q)=1+\left[\frac{2|q|}{Q}\right],
\end{eqnarray*}
for the transition amplitude $\langle k|s,q\rangle $.
In case when $k=nQ/2$ with $n$ integer, hybridization of the 
neighboring bands should be taken into account. This modifies the 
above relations by the following way
 \begin{eqnarray*}
 	c_{nQ/2} =\theta(n)\left[\alpha_{n}C_{n,q_{n}}+
 	\beta_{n}C_{n+1,q_{n}}\right]\\
  +\theta(-n)\left[\beta^{*}_{-n}C_{-n,q_{n}}-
 	\alpha^{*}_{-n}C_{-n+1,q_{n}}\right],\\ 	
 	q_{n}=Q\left(\left\{\frac{n+1}{2}\right\}-
	\frac{1}{2}\right); 
 \end{eqnarray*}
 \begin{eqnarray*}
 	C_{s,q_{s}} =\alpha^{*}_{s}c_{sQ/2}+
 	\beta_{s}c_{-sQ/2},\\
    C_{s+1,q_{s}} =\beta^{*}_{s}c_{sQ/2}-
 	\alpha_{s}c_{-sQ/2}, 
 \end{eqnarray*}
 where $\alpha$, $\beta$ are hybridization coefficients. 
 Corresponding relations between wave functions follow immediately
 from these formulas.\\
  
To write down any of these formulas for a specific array, one should 
add the array index $j$ to the wave function $\psi$, Bloch amplitude 
$u$, coordinate $x$, quasimomentum $q$, and to the periods $a$ and $Q$ 
of the superchain in real and reciprocal space.\\
\section*{Appendix B. Double QCB Spectrum}
Here we obtain analytical expressions for dispersion laws and wave 
functions of QCB. For quasimomenta far from the BZ boundaries, the 
energy spectrum of the first band can be calculated explicitly.  
Assuming that $\omega^2\ll\omega_{js}^2(q_j)$, $s=2,3,4,\ldots$, we 
omit $\omega^{2}$ in all terms in the r.h.s.  of Eq. (\ref{F_j}) 
except  
the first one, $s=1$.  As a result the secular equation (17) reads
\begin{equation}
  \prod\limits_{j=1}^{2}
  \left(
       \frac{\varphi_j^2(q_j)\omega^2}
            {\omega_j^2(q_j)-\omega^2}+
       F_j
  \right)\nonumber
  =\frac{1}{\varepsilon},
\end{equation}
where
$$
  \varphi_j^2(q)=\frac{r_{j}}{a_j}{\phi}_{j1}^2(q),
  \ \ \ \ \
  {\omega_j^2(q)}={\omega_{j1}^2(q_j)}.
$$
The solutions of this equation have the form:
2\begin{eqnarray}
  {\omega}_{{\nu}1{\bf{q}}}^{2} & = &
  \tilde\omega_1^{2}({\bf q})+
        \tilde\omega_2^{2}({\bf q})
  \nonumber\\ &\pm &
       \sqrt{
       \left(
       \tilde\omega_1^{2}({\bf q})-
             \tilde\omega_2^{2}({\bf q})
       \right)^{2}+
       4\varepsilon\varphi^2_{\bf {q}}
       \omega_1^2(q_1)\omega_2^2(q_2)}.
  \label{omega-12}
\end{eqnarray}
Here 
$$
  \varphi_{\bf{q}}={\varphi}_{1}(q_1)
                {\varphi}_{2}(q_2),
$$
$\nu=+,-$ is the branch number, $\tilde\omega_{j}({\bf q})$ is 
determined as
\begin{equation}
	\tilde\omega_{1}^2({\bf q})=
              \omega_{1}^{2}(q_{1})
              \frac{1-\varepsilon{F}_1
                                 (F_2-{\varphi}_{2}^{2}(q_{2}))}
                   {1-\varepsilon(F_1-{\varphi}_{1}^{2}(q_{1}))
                                 (F_2-{\varphi}_{2}^{2}(q_{2}))}
	\label{RenFr}
\end{equation}
for $j=1.$ Expression for $\tilde\omega_{2}^2({\bf q})$ can be
obtained by permutation $1\leftrightarrow 2.$ Parentheses on the
r.h.s. of Eq.  (\ref{RenFr}) describe the contributions to $F_{j}$
from higher bands.  Therefore $\tilde\omega_{j}^2({\bf q})$ is the
$j$-th array frequency renormalized by the interaction with higher
bands.  In principle, contribution of higher bands may turn the
interaction to be strong.  However for specific case of carbon
nanotubes, one stays far from the critical value $\varepsilon_{c}$
(see estimates at the end of subsection \ref{subsec:Approx}). 
Therefore the interaction with higher bands is weak almost in all the
BZ except its boundaries.\\

The resonance line equation modified by interaction with higher bands 
is 
\begin{equation}
	\tilde \omega^{2}_{1}({\bf q})=\tilde \omega^{2}_{2}({\bf q}).
	\nonumber
\end{equation}
Out of this line the branch number is in fact the array number and 
the
renormalized frequencies are frequencies of a boson propagating
along one of the arrays slightly modified by interactions with
the complementary array. In case when 
$\omega_{1}(q_{1})>\omega_{2}(q_{2}),$ one obtains
\begin{equation}
  {\omega}_{+,1{\bf{q}}}^{2}\approx
  \omega_1^2(q_1)
  \left(1-\varepsilon F_2 \varphi_1^2(q_1)\right).
  \label{omega-SecOrd}
\end{equation}
In the opposite case one should replace indices $1\leftrightarrow 2,$
 $-\leftrightarrow+$.\\

Consider the frequency correction in the latter equation in more 
details.  The correction term can be approximately estimated as 
${\omega}_{1}^{2}(q_{1})S(q_{1})$ with
\begin{equation}
   S(q_{1}) =
   \varepsilon{F}_{2}\varphi_1^{2}(q_1) =
   \varepsilon
   \frac{R_0}{a}
   \phi_{11}^{2}(q_1)
   \int d\xi\zeta_2^2(\xi).
    \label{Sq}
\end{equation}
Due to the short-range character of the interaction, the matrix
elements ${\phi}_{11}(q_1)\sim{1}$ vary slowly with the
quasimomentum ${q}_{1}\le{Q}_{1}$. Therefore, the r.h.s. in
Eq.(\ref{Sq}) can be roughly estimated as
\begin{equation}
    S(q_{1})\sim \varepsilon\frac{R_{0}}{a}=0.1
    \frac{R_{0}}{a} \ll 1.
    \label{est1}
\end{equation}
One should also remember that the energy spectrum of nanotube
remains one-dimensional only for frequencies smaller than some
$\omega_{m}.$ Therefore, an external cutoff arises at $s=ak_{m}$
where $k_{m}\sim\omega_{m}/v$. As a results one gets an estimate
\begin{equation}
    S(q_{1})\sim \varepsilon \frac{R_{0}}{a}k_{m}R_{0}.
    \label{est2}
\end{equation}
Hence, one could hope to gain additional power of the small 
interaction radius. However, for nanotubes, $k_{m}$ is of the order of 
$1/R_{0}$ (see Refs.\onlinecite{Ando,Egger}) and both estimates 
coincide.  For quasimomenta close to the BZ center, the coefficient 
$S(q_{1})$ can be calculated exactly.  For exponential form of 
$\zeta(\xi)\propto\exp(-|\xi|)$, one obtains instead of the 
preliminary estimate (\ref{est1}),
$$
     S(0) = 0.14 \frac{R_{0}}{a}.
$$
Thus, the correction term in Eq.(\ref{omega-SecOrd}) is really small.\\

The eigenstates of the system are described by renormalized field 
operators. Within the first band they have the form
\begin{eqnarray}
 \tilde\theta_{11{\bf{q}}} & = &
 \left(
      1-\frac{1}{2}\beta_{1{\bf{q}}}
 \right)
 \left(
      u_{\bf{q}}\theta_{11{\bf{q}}}-
      v_{\bf{q}}\theta_{21{\bf{q}}}
 \right)
 \nonumber\\
 & - &
 \sum\limits_{s=2}^{\infty}
 \left(
      \phi_{1s{\bf{q}}}u_{\bf{q}}\theta_{2s{\bf{q}}}+
      \phi_{2s{\bf{q}}}v_{\bf{q}}\theta_{1s{\bf{q}}}
 \right),
 \label{tilde-teta1}
 \\
 \tilde\theta_{21{\bf{q}}} & = &
 \left(
      1-\frac{1}{2}\beta_{2{\bf{q}}}
 \right)
 \left(
      v_{\bf{q}}\theta_{11{\bf{q}}}+
      u_{\bf{q}}\theta_{21{\bf{q}}}
 \right)
 \nonumber\\
 & - &
 \sum\limits_{s=2}^{\infty}
 \left(
      \phi_{1s{\bf{q}}}v_{\bf{q}}\theta_{2s{\bf{q}}}+
      \phi_{2s{\bf{q}}}u_{\bf{q}}\theta_{1s{\bf{q}}}
 \right).
 \label{tilde-teta2}
\end{eqnarray}
Here the coefficients $u_{\bf{q}}$ and $v_{\bf{q}}$ describe mixing 
between the modes with different array indices, within the first band,
\begin{equation}
 u_{\bf{q}}=
 \sqrt{
      \frac{
            \sqrt{\Delta_{\bf{q}}^{2}+\phi_{1{\bf{q}}}^{2}}+
            \Delta_{\bf{q}}
           }
           {2\sqrt{\Delta_{\bf{q}}^{2}+\phi_{1{\bf{q}}}^{2}}}
      },
 \label{u}
\end{equation}
\begin{equation}
 v_{\bf{q}}=
 \sqrt{
      \frac{
            \sqrt{\Delta_{\bf{q}}^{2}+\phi_{1{\bf{q}}}^{2}}-
            \Delta_{\bf{q}}
           }
           {2\sqrt{\Delta_{\bf{q}}^{2}+\phi_{1{\bf{q}}}^{2}}}
      },
 \label{v}
\end{equation}
and
\begin{eqnarray}
\Delta_{\bf{q}}&=&
\frac{\omega_{21}^{2}(q_2)-\omega_{11}^{2}(q_1)}{2},\nonumber\\
 \phi_{1{\bf{q}}}&=&\sqrt{\varepsilon}
 \varphi_{\bf{q}}\omega_{11}(q_1)\omega_{21}(q_2).
	\label{phi1}
\end{eqnarray}
The parameters $\phi_{1s{\bf{q}}}$, 
$\phi_{2s{\bf{q}}},$ $s=2,3,\ldots,$ in 
Eqs.  (\ref{tilde-teta1}), (\ref{tilde-teta2}) correspond to 
inter-band mixing
\begin{equation}
	\phi_{1s{\bf{q}}}=\sqrt{\varepsilon}
 \frac{r_0}{a}\phi_{11}(q_1)\phi_{2s}(q_2)
 \frac{\omega_{11}(q_1)}{\omega_{2s}(q_2)},\nonumber
\end{equation}
and the coefficients $\beta_{1{\bf q}},$  $\beta_{2{\bf q}},$ take into 
account corrections from the higher bands
 \begin{equation}
 	\beta_{1{\bf{q}}}=
 \sum\limits_{s=2}^{\infty}
     \left(
          \phi_{1s{\bf{q}}}^{2}u_{\bf{q}}^{2}+
          \phi_{2s{\bf{q}}}^{2}v_{\bf{q}}^{2}
     \right).\nonumber
 \end{equation}
Expressions for $\phi_{2s{\bf{q}}}$ and $\beta_{2{\bf{q}}}$ can be 
obtained by permutation ${1}\leftrightarrow{2}$.\\

Equations (\ref{omega-12}), (\ref{tilde-teta1}) and
(\ref{tilde-teta2}) solve the problem of QCB energy spectrum away from
the BZ boundaries.  However, due to smalness of the interaction, the 
general expressions (\ref{tilde-teta1}) and (\ref{tilde-teta2}) can be
simplified.  For quasimomenta far from the resonant coupling line, the
expressions for the renormalized field operators of the first array look
like
\begin{equation}
 {\tilde\theta}_{11{\bf{q}}} =
 \left(1-\frac{1}{2}\tilde\beta_{1{\bf{q}}}\right)
   {\theta}_{11{\bf{q}}}+
 \sum\limits_{s=1}^{\infty}
   {\phi}_{1s{\bf{q}}}
   {\theta}_{2s{\bf{q}}},\nonumber
\end{equation}
where
$$
 \phi_{11{\bf{q}}}=\sqrt{\varepsilon}
 \frac{r_0}{a}\phi_{11}(q_1)\phi_{21}(q_2)
 \frac{\omega_{11}(q_1)\omega_{21}(q_2)}
      {\omega_{21}^2(q_2)-\omega_{11}^2(q_1)},
$$
and 
\begin{equation}
   \tilde\beta_{1{\bf q}}=\sum\limits_{s=1}^{\infty}
   {\phi}_{1s{\bf{q}}}^{2}.\nonumber
\end{equation}
The corresponding formulas for the second array are obtained by
replacing $1s\rightarrow 2s$.\\
  
Another simplification is made for modes with quasi--momenta on the
resonance line.  Consider for simplicity a square QCB (in this case BZ
coincides with the elementary cell of the reciprocal lattice, and the
resonance line coincides with the BZ diagonal $OC$ in
Fig.\ref{BZ2}) and assume that ${\bf q}$ is not too close to the BZ
corner $C$.  The initial frequencies of modes belonging to the same
band coincide,
\begin{equation}
	{\omega}_{1s{\bf q}}= {\omega}_{2s{\bf
q}}\equiv {\omega}_{s{\bf q}}.\nonumber
\end{equation}
Therefore renormalization strongly mixes the initial variables
\begin{eqnarray*}
{\tilde\theta}_{gs{\bf{q}}} & = & \frac{1}{\sqrt{2}}
  \left(
       1-\frac{1}{2}
       \beta_{s{\bf q}}
  \right)
  ({\theta}_{2s{\bf{q}}}+
       {\theta}_{1s{\bf{q}}})
\\
& - &
   \frac{1}{\sqrt{2}}
   \sum\limits_{s'\neq s}  
      ({\phi}_{s's{\bf{q}}}
      {\theta}_{1s'{\bf{q}}}-
      {\phi}_{ss'{\bf{q}}}
      {\theta}_{2s'{\bf{q}}}),
\end{eqnarray*}
\begin{eqnarray*}
{\tilde\theta}_{usþ{\bf{q}}} & = & \frac{1}{\sqrt{2}}
  \left(
       1-\frac{1}{2}
       {\beta}_{s{\bf q}}
  \right)
  ({\theta}_{2s{\bf{q}}}
      -{\theta}_{1s{\bf{q}}})\\
& - & 
   \frac{1}{\sqrt{2}}
   \sum\limits_{s'\neq s}
   ({\phi}_{s's{\bf{q}}}
      {\theta}_{1s'{\bf{q}}}+
      {\phi}_{ss'{\bf{q}}}
      {\theta}_{2s'{\bf{q}}}),
\end{eqnarray*}
and the corresponding eigenfrequencies are shifted from their bare
values
\begin{eqnarray*}
{\omega}_{gs{\bf{q}}}^{2} & \approx &
      {\omega}_{s{\bf{q}}}^{2}
       \left(
            1+
            {\phi}_{1s2s{\bf{q}}}\right),
  \\
{\omega}_{us{\bf{q}}}^{2} & \approx &
      {\omega}_{s{\bf{q}}}^{2}
       \left(
            1-
            {\phi}_{1s2s{\bf{q}}}\right).
\end{eqnarray*}

For the first band $s=1$ these formulas look like
\begin{equation}
  {\omega}_{\pm ,1{\bf{q}}}^{2}\approx
  \omega_1^2(q_1)
  \left(1\pm\sqrt{\varepsilon}\varphi_{\bf{q}}\right).\nonumber
\end{equation}
Note that in the resonance case the splitting of the degenerate modes 
is of the order of $\sqrt{\varepsilon}$ that essentially exceeds the 
shift 
of eigenfrequencies in the non-resonant case (\ref{omega-SecOrd}).\\

The interband mixing becomes significant near the BZ boundaries.  
Not very close to the crossing points of these boundaries with the 
resonant lines, this mixing is accounted for by a standard way.  As a 
result we find that the interband hybridization gap for the bosons 
propagating along the first array can be estimated as
\begin{eqnarray*}
  {\Delta\omega}_{12}\sim
 {vQ}\varepsilon \frac{r_{0}}{a}.
\end{eqnarray*}
Similar gaps exist near the boundary of the BZ for each pair of odd and 
next even energy bands, as well as for each even and next odd band near 
the lines $q_1=0$ or $q_2=0$.  The energy gap between the $s$-th and 
$(s+1)$-th bands is estimated as
\begin{eqnarray*}
  {\Delta\omega}_{s,s+1}\sim
  {vQ}\varepsilon\frac{r_{0}}{a}{\it o}(s^{-1}).
\end{eqnarray*}
For large enough band number $s,$ the interaction is 
effectively suppressed, 
${\phi}_{1s2s'}\to{0},$ and the gaps vanish.\\

The spectral behavior in the vicinity of the crossing points of a 
resonance line and the BZ boundary needs more detailed calculations.  
Nevertheless it can also be analyzed in a similar way.  The results of 
such an alalysis are discussed in subsubsection \ref{subsubsec:Square}.\\

\section*{Appendix C. {\it ac} conductivity}
For interacting wires, where ${\phi}_{js}({q}_j)\neq{0},$ the 
correlator (\ref{CurrCorr}) may be easily calculated after 
diagonalization of the Hamiltonian (\ref{TotHam2}) by means of the 
transformations (\ref{tilde-teta1}) and (\ref{tilde-teta2}).  As a 
result, one has:
\begin{eqnarray*}
\left\langle\left[
                 {J}_{11{ \bf{q}}}(t),
                 {J}_{11{\bf{q}}}^{\dag}(0)
  \right]\right\rangle = \\
      -2ivg\left(u_{\bf{q}}^{2}
      {\omega}_{+,1{\bf{q}}}
      \sin({\omega}_{+,1{\bf{q}}}t)
      +v_{\bf{q}}^{2}
      {\omega}_{-,1{\bf{q}}}
      \sin({\omega}_{-,1{\bf{q}}}t)\right),
\end{eqnarray*}
\begin{eqnarray*}
  &
\left\langle\left[
                 {J}_{11{ \bf{q}}}(t),
                 {J}_{21{\bf{q}}}^{\dag}(0)
\right]\right\rangle  =
  -2ivgu_{\bf{q}}v_{\bf{q}}
  & \\
  & \times \left(
     {\omega}_{-,1{\bf{q}}}
     \sin({\omega}_{-,1{\bf{q}}}t)-
     {\omega}_{+,1{\bf{q}}}
     \sin({\omega}_{+,1{\bf{q}}}t)
 \right), &
\end{eqnarray*}
where $u_{\bf{q}}$ and $v_{\bf{q}}$ are defined in Eqs.(\ref{u}),
(\ref{v}). Then, for the optical absorption ${\sigma}'$ one obtains
\begin{eqnarray}
& {\sigma}'_{11}({\bf{q}},\omega)  =
    {\pi}{v}{g}
    \Bigl[
         u_{\bf{q}}^{2}
         \delta
         \left(
              {\omega}-
              {\tilde\omega}_{+,1{\bf{q}}}
         \right)+
         v_{\bf{q}}^{2}
         \delta
         \left(
              {\omega}-
              {\tilde\omega}_{-,1{\bf{q}}}
         \right)
    \Bigr]
    \label{opt_l}
\end{eqnarray}
\begin{equation}
   {\sigma}'_{12}({\bf{q}},\omega)  =
    {\pi}{v}{g}
    u_{\bf{q}}v_{\bf{q}}
    \Bigl[
         \delta
         \left(
              {\omega}-
              {\tilde\omega}_{-,1{\bf{q}}}
         \right)-
         \delta
         \left(
              {\omega}-
              {\tilde\omega}_{+,1{\bf{q}}}
         \right)
    \Bigr].
    \label{opt_t}
\end{equation}
For quasimomentum ${\bf{q}}$ away from the resonant coupling line,
$u_{\bf{q}}^{2}\approx{1}$ and $v_{\bf{q}}^{2}\sim\phi_{\bf{q}}^{2}$
for $\Delta_{\bf{q}}>0$ ($v_{\bf{q}}^{2}\approx{1}$ and
$u_{\bf{q}}^{2}\sim\phi_{\bf{q}}^{2}$ for $\Delta_{\bf{q}}<0$).  Then
the longitudinal optical absorption (\ref{opt_l}) (i.e. the absorption
within a given set of wires) has its main peak at the frequency
${\omega}_{+,1{\bf{q}}}\approx{v\vert{q_1}\vert}$ for
$\Delta_{\bf{q}}>0$ (or
${\omega}_{-,1{\bf{q}}}\approx{v\vert{q_1}\vert}$ for
$\Delta_{\bf{q}}<0$), corresponding to the first band of the pertinent
array, and an additional weak peak at the frequency
${\omega}_{-,1{\bf{q}}}\approx{v\vert{q_2}\vert}$, corresponding to
the first band of a complementary array.  It contains also a set of
weak peaks at frequencies $\omega_{2,s{\bf{q}}}\approx [s/2]vQ$
($s=2,3,\ldots$) corresponding to the contribution from higher bands
of the complementary array (in Eq.(\ref{opt_l}) these peaks are
omitted).  At the same time, a second observable becomes relevant,
namely, the transverse optical absorption (\ref{opt_t}).  It is
proportional to the (small) interaction strength and has two peaks at
frequencies ${\omega}_{+,1{\bf{q}}}$ and ${\omega}_{-,1{\bf{q}}}$ in
the first bands of both sets of wires.\\

If the quasimomentum ${\bf{q}}$ belongs to the
resonant coupling line $\Delta_{\bf{q}}=0$, then 
$u_{\bf{q}}^{2}=v_{\bf{q}}^{2}=1/2$. In this case the longitudinal 
optical
absorption (\ref{opt_l}) has a split double peak at frequencies
${\omega}_{+,1{\bf{q}}}$ and ${\omega}_{-,1{\bf{q}}}$,
instead of a single main peak. The transverse optical absorption
(\ref{opt_t}), similarly to the non-resonant case (\ref{opt_t}),
has a split double peak at frequencies
${\omega}_{+,1{\bf{q}}}$ and ${\omega}_{-,1{\bf{q}}}$, but 
its amlitude is now of the order of unity.
For $\left\vert{\bf{q}}\right\vert\to 0$ Eq.(\ref{opt_l}) reduces
to that for an array of noninteracting wires (\ref{Drude_peak}),
and the transverse optical conductivity (\ref{opt_t}) vanishes.\\

The imaginary part of the {\it ac} conductivity
${\sigma}''_{jj'}({\bf{q}},\omega)$ is calculated within the same
approach. Its longitudinal component equals
\begin{eqnarray*}
&{\sigma}''_{11}({\bf{q}},\omega)=
            \displaystyle{\frac{2vg}{\omega}
            \left[
                 \frac{u_{\bf{q}}^{2}
                       {\omega}_{+,1{\bf{q}}}^{2}}
                      {{\omega}_{+,1{\bf{q}}}^{2}-
                       \omega^2}+
            \frac{v_{\bf{q}}^{2}
                      {\omega}_{-,1{\bf{q}}}^{2}}
                      {{\omega}_{-,1{\bf{q}}}^{2}-
                       \omega^2}
            \right]}.
\end{eqnarray*}
Beside the standard pole at zero frequency, the imaginary part has
poles at the resonance frequencies $\omega_{+,1{\bf q}}$,
$\omega_{-,1{\bf q}}$, and an additional series of high band
satellites (omitted here).  For quasimomenta far from the resonant
lines, only the first pole is well pronounced while amplitude of the
second one as well as amplitudes of all other sattelites is small.  At
the resonant lines, amplitudes of both poles mentioned above are equal. 
The corresponding expression for ${\sigma}'_{22}({\bf{q}},\omega)$ can be
obtained by replacement $1\leftrightarrow 2$.  \\

The transverse component of the
imaginary part of the {\it ac} conductivity has the form:
\begin{eqnarray*}
{\sigma}'_{12}({\bf{q}},\omega)=
        \displaystyle{
        \frac{2vg}{\omega}u_{{\bf q}}v_{{\bf q}}
        \left[
             \frac{{\omega}_{-,1{\bf{q}}}^{2}}
                  {{\omega}^{2}-{\omega}_{-,1{\bf{q}}}^{2}}-
             \frac{{\omega}_{+,1{\bf{q}}}^{2}}
                  {{\omega}^{2}-{\omega}_{+,1{\bf{q}}}^{2}}
        \right]}.
\end{eqnarray*}
It always contains two poles and vanishes for noninteracting wires. 
For quasimomenta far from the resonance lines the transverse component is 
small while at these lines its amplitude is of the order of unity.\\
\section*{Appendix D. Triple QCB Spectrum}
To diagonalize the Hamiltonian(\ref{H}), we write down equations of motion
\begin{eqnarray}
 \left[
      \omega_s^2(q_{j})-\omega^2
 \right]\theta_{js{\bf{q}}}\nonumber\\
 +\sqrt{\varepsilon}\phi_s(q_j)\omega_s(q_j)
 \frac{r_0}{a}
 \sum\limits_{s'}
 \phi_{s'}(q_3)\omega_{s'}(q_3)
 \theta_{3s'{\bf{q}}}
 &=& 0,
 \label{EqMotion1-2}
 \\
 \left[
      \omega_s^2(q_{3})-\omega^2
 \right]\theta_{3s{\bf{q}}}\nonumber\\
 +\sqrt{\varepsilon}\phi_s(q_3)\omega_s(q_3)
 \frac{r_0}{a}
 \sum\limits_{j,s'}
 \phi_{s'}(q_j)\omega_{s'}(q_j)
 \theta_{js'{\bf{q}}}
 &=& 0.
 \label{EqMotion3}
\end{eqnarray}
Here $j=1,2$, and $\varepsilon$ is defined by Eq.(\ref{epsilon1}).
The solutions of the set of equations
(\ref{EqMotion1-2}) - (\ref{EqMotion3}) have the form:
\begin{eqnarray}
 \theta_{js{\bf{q}}} &=& A_j
 \frac{\phi_s(q_j)\omega_s(q_j)}
      {\omega_s^2(q_j)-\omega^2}.
 \ \ \ \ \ j=1,2,3,\nonumber
\end{eqnarray}
Substituting
this equation into Eqs.(\ref{EqMotion1-2}) and
(\ref{EqMotion3}), we have three equations for constants $A_j$:
\begin{eqnarray*}
 A_1+A_3{\sqrt{\varepsilon}}F_{q_3}(\omega^2) &=& 0,\\
 A_2+A_3{\sqrt{\varepsilon}}F_{q_3}(\omega^2) &=& 0,\\
 A_3+\sum\limits_{j=1,2}A_j{\sqrt{\varepsilon}}F_{q_j}(\omega^2) &=& 0,
\end{eqnarray*}
where
\begin{equation}
 F_q(\omega^2)=\frac{r_0}{a}\sum\limits_{s}
 \frac{\phi_s^2(q)\omega_s^2(q)}{\omega_s^2(q)-\omega^2}.\nonumber
\end{equation}
Dispersion relations can be obtained from the solvability condition 
for this set of equations
\begin{equation}
 \varepsilon F_{q_3}(\omega^2)
 \left(
      F_{q_1}(\omega^2)+
      F_{q_2}(\omega^2)
 \right)=1.\nonumber
\end{equation}
The function $F_{q_{s}}(\omega^2)$ has a set of poles at
$\omega^2=\omega_{s}^{2}(q)$, $s=1,2,3,\ldots$ .  For 
$\omega^{2}<\omega_{s}^{2}(q)$, i.e. within the interval
$[0,\omega_{1}^{2}(q)]$,  $F_{q_{s}}(\omega^2)$ is positive 
increasing function.
Its minimal value $F$ on the interval is reached at $\omega^2=0$
and does not depend on quasi-momentum $q$
\begin{equation}
 F_q(0) =\frac{r_0}{a}\sum\limits_{s}
 \phi_{s}^{2}(q)=\int d\xi \zeta_j^2(\xi)\equiv F.\nonumber
\end{equation}
If the parameter $\varepsilon\equiv\eta^{2}$ is smaller than the critical 
 value
\begin{equation}
  \varepsilon_c=
  \frac{1}{2F^{2}},\nonumber
\end{equation}
then all solutions $\omega^{2}$ of the characteristic equation are 
positive.  When $\varepsilon$ increases, the lowest QCB mode softens 
and its square frequency vanishes \textit{in whole BZ} at 
$\varepsilon=\varepsilon_{c}$.  For exponential model 
$\zeta(\xi)=\exp(-|\xi|),$ one obtains $\varepsilon_c\approx 1$.\\
\section*{Appendix E. Triple Rabi Oscillations}
The point $C(Q/3,2Q/3)$  of the BZ is the point of
three-fold degeneracy, 
$$
q_{1}=q_{3}=-q_{2}+Q=\frac{Q}{3},
$$
$$
  \omega_{11}(Q/3)=\omega_{21}(2Q/3)=\omega_{31}(Q/3)\equiv\omega_0.
$$
Equations of motion at this point in the resonance approximation read
\begin{eqnarray*}
  &&\left[\frac{d^2}{dt^2}+\omega_{0}^{2}\right]
  \theta_{1}+
  \sqrt{\varepsilon}\phi^2\omega_{0}^2\theta_{3} = 0,\\
  &&\left[\frac{d^2}{dt^2}+\omega_{0}^{2}\right]
  \theta_{2}+
  \sqrt{\varepsilon}\phi^2\omega_{0}^2\theta_{3} = 0,\\
  &&\left[\frac{d^2}{dt^2}+\omega_{0}^{2}\right]
  \theta_{3}+
  \sqrt{\varepsilon}\phi^2\omega_{0}^2
  \left(\theta_{1}+\theta_{2}\right) = 0,
\end{eqnarray*}
where $\theta_{j}\equiv\theta_{j{\bf q}}.$
General solution of this system looks as
\begin{eqnarray*}
	\left(\begin{array}{c}\theta_{1}(t)\\
	\theta_{2}(t)\\
	\theta_{3}(t)\end{array}\right)=
	\theta_{0}\left(\begin{array}{c}
	1\\
	-1\\
	0\end{array}\right)e^{i\omega_{0}t}+\ \ \ \ \ \ \ \ \ \ \ \ 
	\\
	\theta_{+}\left(\begin{array}{c}
	1\\
	1\\
	\sqrt{2}\end{array}\right)e^{i\omega_{+}t}+
	\theta_{-}\left(\begin{array}{c}
	1\\
	1\\
	-\sqrt{2}\end{array}\right)e^{i\omega_{-}t},
\end{eqnarray*}
where one of the eigenfrequencies coincides with $\omega_{0},$ while
the two others are
\begin{equation}
	\omega_{\pm}=\sqrt{1\pm\sqrt{2}\phi^{2}},\nonumber
\end{equation}
and $\theta_{0,\pm}$ are the corresponding amplitudes.\\

Choosing initial conditions
\begin{eqnarray*}
  &&{\theta}_{1}(0)=i\theta_0, \ \ \ \ \ \ \
  {\dot{\theta}}_{1}(0)=\omega_0\theta_0,\\
  &&{\theta}_{2}(0)=0,
  \ \ \ \ \ \ \ \
  {\dot{\theta}}_{2}(0)=0,\\
  &&{\theta}_{3}(0)=0,
  \ \ \ \ \ \ \ \
  {\dot{\theta}}_{3}(0)=0,
\end{eqnarray*}
we obtain for the field amplitudes at the coordinate origin
\begin{eqnarray*}
  \theta_1(0,0;t) &=& \frac{\theta_0}{4}
  \left[
       \frac{\omega_0}{\omega_+}\sin(\omega_+t)+
       \frac{\omega_0}{\omega_-}\sin(\omega_-t)
  \right]
  \\&&+
  \frac{\theta_0}{2}\sin(\omega_0t),
  \nonumber\\
  \theta_2(0,0;t) &=& \frac{\theta_0}{4}
  \left[
       \frac{\omega_0}{\omega_+}\sin(\omega_+t)+
       \frac{\omega_0}{\omega_-}\sin(\omega_-t)
  \right]
  \\&&-
  \frac{\theta_0}{2}\sin(\omega_0t),
  \nonumber\\
  \theta_3(0,0;t) &=& \frac{\theta_0}{2\sqrt{2}}
  \left[
       \frac{\omega_0}{\omega_+}\sin(\omega_+t)-
       \frac{\omega_0}{\omega_-}\sin(\omega_-t)
  \right].
\end{eqnarray*}
In the limiting case $\varepsilon\ll 1$ these formulas coincide
with Eqs.(\ref{sol3}) in subsection {\bf III.C}.

\end{document}